\documentclass[preprint,12pt]{elsarticle}

\usepackage{amsmath,amssymb,amsfonts}
\usepackage{graphicx}
\usepackage{textcomp}
\usepackage{xcolor}

\usepackage{hyperref}
\usepackage{amstext}
\usepackage{graphics}
\usepackage{url}
\usepackage{subfigure}
\usepackage{latexsym}
\usepackage{longtable}
\usepackage{multirow}
\usepackage{verbatim}
\usepackage{xspace}
\makeatletter
\newif\if@restonecol
\makeatother

\usepackage{enumitem}

\usepackage{pifont}
\usepackage{url}
\usepackage[ruled,vlined,linesnumbered]{algorithm2e}
\usepackage{algpseudocode}{\fontsize{6.5pt}{\baselineskip}\selectfont}
\newcommand{\vsmall}{\fontsize{7.2pt}{\baselineskip}\selectfont}
\newcommand{\spara}[1]{\smallskip\noindent{\bf #1}}
\usepackage[font=small]{caption}

\newtheorem{definition}{Definition}
\newtheorem{example}{Example}
\SetKwFor{Loop}{loop}{}{endloop}

\journal{Knowledge-based systems}

\begin{document}

\begin{frontmatter}



\title{Efficiently Embedding Dynamic Knowledge Graphs}


\author[seu]{Tianxing Wu\corref{cor1}}
\ead{tianxingwu@seu.edu.cn}
\author[abu]{Arijit Khan}
\author[nus]{Melvin YONG}
\author[seu]{Guilin Qi}
\author[seu]{Meng Wang}

\cortext[cor1]{Corresponding author}
\address[seu]{Southeast University, China}
\address[abu]{Aalborg University, Denmark}
\address[nus]{National University of Singapore, Singapore}

\begin{abstract}
Knowledge graph (KG) embedding encodes the entities and relations from a KG into low-dimensional vector spaces to support various applications such as KG completion, question answering, and recommender systems. In real world, knowledge graphs (KGs) are dynamic and evolve over time with addition or deletion of triples. However, most existing models focus on embedding static KGs while neglecting dynamics. To adapt to the changes in a KG, these models need to be retrained on the whole KG with a high time cost. In this paper, to tackle the aforementioned problem, we propose a new context-aware Dynamic Knowledge Graph Embedding (DKGE) method which supports the embedding learning in an online fashion. DKGE introduces two different representations (i.e., knowledge embedding and contextual element embedding) for each entity and each relation, in the joint modeling of entities and relations as well as their contexts, by employing two attentive graph convolutional networks, a gate strategy, and translation operations. This effectively helps limit the impacts of a KG update in certain regions, not in the entire graph, so that DKGE can rapidly acquire the updated KG embedding by a proposed online learning algorithm. Furthermore, DKGE can also learn KG embedding from scratch. Experiments on the tasks of link prediction and question answering in a dynamic environment demonstrate the effectiveness and efficiency of DKGE.
\end{abstract}




\end{frontmatter}

\section{Introduction}\label{sec:intro}
Knowledge graphs (KGs) such as DBpedia~\cite{lehmann2015dbpedia}, YAGO~\cite{mahdisoltani2013yago3}, and Freebase~\cite{bollacker2008freebase}, have been built to benefit many intelligent applications, e.g., semantic search, question answering, and recommender systems. These KGs are multi-relational graphs describing entities and their relations in the form of triples. A triple is often denoted as (\emph{head entity}, \emph{relation}, \emph{tail entity}) (i.e., $(h,r,t)$) to indicate that two entities are connected by a specific relation, e.g., (\emph{Barack Obama}, \emph{Party}, \emph{Democratic Party}). Recently, techniques of knowledge graph (KG) embedding~\cite{wang2017knowledge} have received considerable attention, as they can learn the representations (i.e., embeddings) of entities and relations in low-dimensional vector spaces, and these embeddings can be used as features to support link prediction, entity classification, and question answering, among many others.

In the real world, KGs are dynamic and always change over time. For example, DBpedia extracts the update stream of Wikipedia each day to keep the KG up-to-date~\cite{hellmann2009dbpedia}. The Amazon product KG needs to be updated quite frequently because there are many new products every day~\cite{dong2018challenges}. However, most existing KG embedding models~\cite{nickel2011three,bordes2013translating,wang2014knowledge,yang2015embedding,trouillon2016complex,feng2016gake,conve2018,sun2018rotate,li2022transo} focus on embedding static KGs while neglecting dynamic updates. To adapt to the changes in a KG, these models need to be retrained on the whole KG with a high time cost, but it is unacceptable when the KG has a high update frequency (e.g., once per day). Thus, \textbf{how to embed dynamic KGs in an online manner} is an important problem to solve.

Although many methods on dynamic graph embedding~\cite{hamilton2017inductive,ma2018depthlgp,zhu2018high,li2017attributed,du2018dynamic,trivedi2019dyrep,liu2020dynamic,liu2021motif,barros2021survey} supporting the online learning of node embeddings have emerged, these methods cannot be applied in dynamic KG embedding, because they only learn node embeddings based on structural proximities without considering relation semantics on edges, but KG embedding needs to learn not only node (entity) embeddings but also relation embeddings, and preserves relational constraints between entities. In addition, some models~\cite{dasgupta2018hyte,jiang2016towards,trivedi2017know,goel2020diachronic,jin2020recurrent,li2021temporal} on temporal KG embedding also work on dynamic KGs, but their target is to mine evolving knowledge from multiple given snapshots of a KG to better perform link and time prediction. In other words, these models conduct only offline embedding learning, but when faced with KG updates, they also need to be retrained on the whole KG, so they cannot embed dynamic KGs with high efficiency.

\begin{figure}[t]
\centering
\subfigure[] {\includegraphics[width=0.2907\textwidth]
{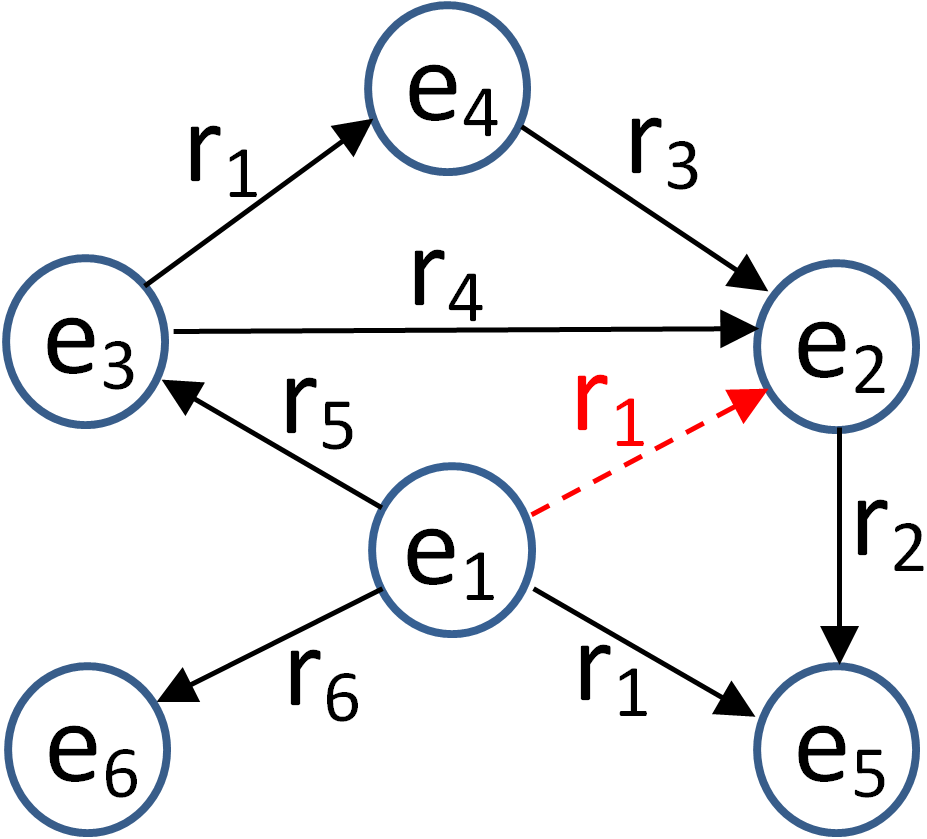}}
\hspace{0.45in}
\subfigure[] {\includegraphics[width=0.49761\textwidth]
{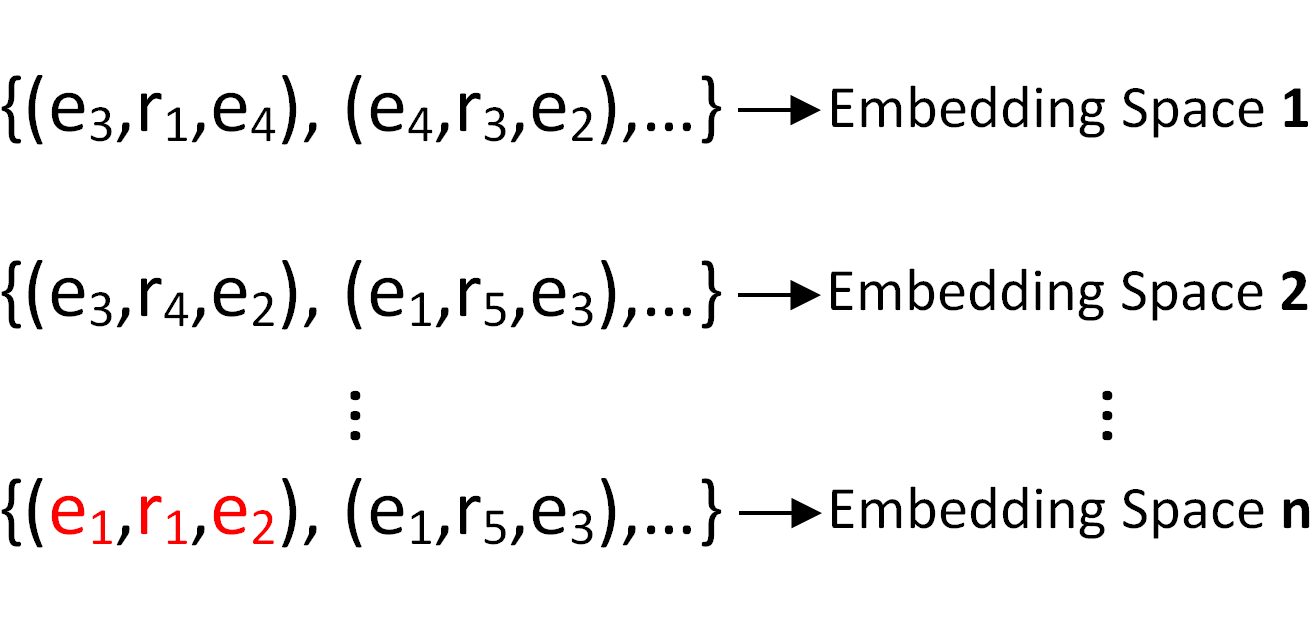}}
\caption{(a) A KG $\mathcal{G}$ does not have the relation $r_1$ between entities $e_1$ and $e_2$ at time step $\mathcal{T}$, and we add a triple $(e_1,r_1,e_2)$ at time step $\mathcal{T}+1$. (b) An illustration of using puTransE~\cite{tay2017non} on $\mathcal{G}$.}
\label{fig:exp_dy}
\end{figure}

The main reason why most KG embedding models are incapable of online embedding learning is: when a KG has an update with addition and deletion of triples, if we revise the representations of some entities and relations to adapt to the updated KG, such revisions will probably spread to the entire graph by correlations among entities and relations. For example, suppose we embed a KG $\mathcal{G}$ (shown in Figure~\ref{fig:exp_dy}(a)) using TransE~\cite{bordes2013translating}, which constrains $\boldsymbol{h}+\boldsymbol{r}\approx\boldsymbol{t}$ (bold characters denote vectors) on each triple $(h,r,t)$, and after adding a new triple $(e_1,r_1,e_2)$ into $\mathcal{G}$,
where $e_1$, $e_2$ are existing entities and $r_1$ is an existing relation in the earlier version of $\mathcal{G}$, we now need to optimize $\boldsymbol{e_1}+\boldsymbol{r_1}\approx\boldsymbol{e_2}$. Regardless of which element in $(e_1,r_1,e_2)$ we choose to revise its representation, it will break the constraint $\boldsymbol{h}+\boldsymbol{r}\approx\boldsymbol{t}$ for other triples containing our chosen element, so it may cause a chain reaction of revisions on the embeddings of entities and relations in the entire graph.

The only existing work that supports online KG embedding learning is puTransE~\cite{tay2017non}. As illustrated in Figure~\ref{fig:exp_dy}(b), puTransE first splits the KG $\mathcal{G}$ into different small sets of triples (a triple may exist in multiple sets), each of which is utilized to train an embedding space, and then only selects the maximum energy score (i.e., $-\|\boldsymbol{h}+\boldsymbol{r}-\boldsymbol{t}\|$ where $\|\cdot\|$ is the $\ell_{1}$ or $\ell_{2}$ norm) of each triple across these embedding spaces for link prediction. When facing a KG update, to support online learning, puTransE directly trains new embedding spaces with small sets of triples containing newly added triples, and deletes existing spaces containing deleted triples. However, puTransE has two major problems which lower the quality of generated embeddings as follows:
\begin{itemize}
  \item \textbf{Problem 1.} puTransE learns embeddings of entities and relations from local parts of a KG, so it avoids retraining on the entire graph when the KG has an update, but this cannot preserve the global structure information of the KG in the learnt embeddings.
  \item \textbf{Problem 2.} puTransE leverages the scoring function of TransE~\cite{bordes2013translating} to compute the energy scores of each triple, which cannot work well to model 1-to-N, N-to-1, and N-to-N relations. Taking a 1-to-N relation $r_1$ as an example, using puTransE to learn embeddings of the entities and relations in triples $(e_1,r_1,e_2)$ and $(e_1,r_1,e_5)$ (see Figure~\ref{fig:exp_dy}(a)) in a space will cause $\boldsymbol{e_2}\approx\boldsymbol{e_5}$.
\end{itemize}

In this paper, we study how to efficiently learn high-quality embeddings of entities and relations in dynamic KGs. Based on the above analyses, we find that it is non-trivial and cannot be well solved by existing KG embedding models. This motivates us to propose a new method which can learn KG embedding from scratch, support online embedding learning, and address the problems of puTransE. To this aim, we devise a novel context-aware \textbf{D}ynamic \textbf{K}nowledge \textbf{G}raph \textbf{E}mbedding method, called \textbf{DKGE}, which can embed dynamic KGs with high effectiveness and efficiency.

\begin{figure}[t]
  \centering
  \includegraphics[width=0.85\textwidth]{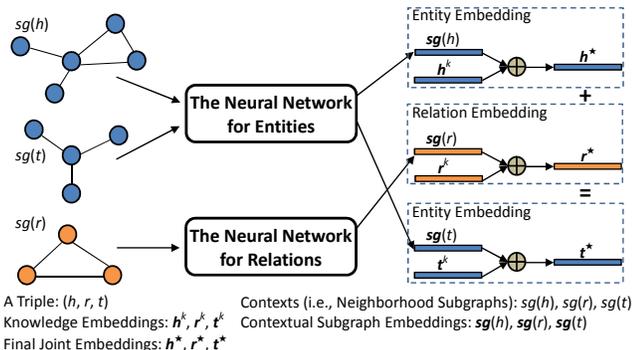}
\caption{The architecture of learning embeddings in DKGE.}
\label{fig:score_f}
\end{figure}

For each triple $(h,r,t)$, unlike puTransE that only uses an individual representation for each entity or relation in the scoring function, DKGE incorporates the contextual information into a joint embedding of each entity (denoted as $\boldsymbol{h^\star}$ and $\boldsymbol{t^\star}$) or relation (denoted as $\boldsymbol{r^\star}$) for the translation operation, i.e., $\boldsymbol{h^\star}+\boldsymbol{r^\star}\approx\boldsymbol{t^\star}$. The context of an entity consists of itself and its neighbor entities. The context of a relation is composed of itself and the relation paths connecting the same entity pairs. These contexts are represented as neighborhood subgraphs. As shown in Figure~\ref{fig:score_f}, the joint embedding of each entity ($\boldsymbol{h^\star}$ or $\boldsymbol{t^\star}$) or relation ($\boldsymbol{r^\star}$) is formed by combining the embedding of itself (called knowledge embedding, i.e., $\boldsymbol{h}^k$, $\boldsymbol{t}^k$, or $\boldsymbol{r}^k$) and the embedding of its context (called contextual subgraph embedding, i.e., $\boldsymbol{sg}(h)$, $\boldsymbol{sg}(t)$, or $\boldsymbol{sg}(r)$) through a gate strategy~\cite{xu2017kg}. Contextual subgraph embeddings of entities and relations are computed by two neural networks, called attentive graph convolutional networks (AGCNs), respectively. The above techniques enable DKGE to learn KG embedding from scratch and \textbf{well model 1-to-N, N-to-1, and N-to-N relations, thereby solving the problem 2 of puTransE}. For example, when modeling triples $(e_1,r_1,e_2)$ and $(e_1,r_1,e_5)$ in Figure~\ref{fig:exp_dy}(a), $\boldsymbol{e_2}\neq\boldsymbol{e_5}$ as long as their contextual subgraph embeddings are different.

To support online learning, DKGE assigns two different representations to each entity or relation. When an entity (or a relation) denotes itself, we use a representation called knowledge embedding; when it denotes a part of the context of other entities (or relations), we use another representation called contextual element embedding. Contextual element embeddings are combined to form contextual subgraph embeddings by using an attentive graph convolutional network (AGCN). Under this setting, we propose an online learning algorithm to incrementally learn KG embedding. In this algorithm, based on the idea of inductive learning, we keep all learnt parameters in AGCNs and the gate strategy unchanged, and contextual element embeddings of existing entities and relations unchanged. After a KG update, there will exist many triples in which the contexts of all entities and relations are unchanged, so their contextual subgraph embeddings are unchanged. Thus, with existing knowledge embeddings of such entities and relations, these triples already hold $\boldsymbol{h^\star}+\boldsymbol{r^\star}\approx\boldsymbol{t^\star}$, so we also keep the knowledge embeddings of existing entities and relations unchanged as long as their contexts are unchanged. In this way, we only need to learn knowledge embeddings and contextual element embeddings of emerging entities and relations, as well as knowledge embeddings of existing entities and relations with changed contexts. This greatly \textbf{reduces the number of triples which need to be retrained while preserving} $\boldsymbol{h^\star}+\boldsymbol{r^\star}\approx\boldsymbol{t^\star}$ \textbf{on the whole KG}. Thus, our algorithm can effectively perform online learning with high efficiency and \textbf{solve the problem 1 of puTransE}.

In the experiments, we first evaluate DKGE on link prediction in a dynamic environment. Compared with state-of-the-art static KG embedding methods, DKGE has comparable effectiveness in different evaluation metrics, and much better efficiency in online learning since the baselines need to be retrained on the whole KG. When comparing with the dynamic KG embedding baseline, i.e., puTransE, DKGE significantly outperforms it in both effectiveness and efficiency. We also conduct case studies on question answering in a dynamic environment to show that DKGE can help obtain accurate answers without writing structured queries in query languages~\cite{liu2022kgvql}.

\spara{Contributions.} The main contributions of this paper are summarized as follows:
\begin{itemize}
  \item We define the problem of embedding dynamic KGs, which is divided into two sub-problems: learning from scratch and online learning (Section~\ref{sec:prodef}).
  \item We propose a new context-aware dynamic KG embedding method DKGE, which can not only learn KG embedding from scratch (Section~\ref{sec:lfs}), but also incrementally learn KG embedding by using an online learning algorithm with high efficiency (Section~\ref{sec:ol}). DKGE solves the problems of puTransE, which is the only existing model supporting online KG embedding learning.
  \item We present a unified solution to encode contexts of entities and relations based on an AGCN model, which can select the most important information from the context of the given entity or relation (Section~\ref{sec:lfs}).
  \item We conduct comprehensive experiments on real-world data management applications, including link prediction and question answering (QA), in a dynamic environment. QA with KG embedding techniques can query the triples which are not in the KG, while classical strategies using structured queries in query languages cannot return any result. The evaluation results show the effectiveness and efficiency of our method DKGE (Section~\ref{sec:exp}).
\end{itemize}

DKGE substantially extends our short conference paper~\cite{fei2021online} in three aspects. The first one is that we conduct more detailed analysis on the differences between dynamic KG embedding, static KG embedding, and dynamic graph embedding, and explicitly point out the problems of current dynamic KG embedding. The second one is that DKGE designs a new attentive graph convolutional network to model contexts of entity and relations, and this new method can characterize different weights of the elements in contexts. The last one is that we perform more comprehensive experiments on four real-world datasets to demonstrate the effectiveness, efficiency, and scalability of DKGE in not only link prediction, but also question answering.

The rest of this paper is organized as follows. Section~\ref{sec:rewo} introduces related work. Section~\ref{sec:prodef} defines our research problem. Section~\ref{sec:lfs} and Section~\ref{sec:ol} present the details of learning from scratch and online learning, respectively, in DKGE. Section~\ref{sec:exp} gives the experimental results and finally we conclude in the last section.

\section{Problem Definition}\label{sec:prodef}
In this section, we define the problem of embedding dynamic KGs as two sub-problems, i.e., learning from scratch and online learning.
Let a KG $\mathcal{G}^{\mathcal{T}}=\{(h,r,t)\}\subseteq \mathcal{E}\times\mathcal{R}\times\mathcal{E}$, where $\{(h,r,t)\}$ represents a set of triples, $h$ means a head entity, $r$ is a relation, $t$ is a tail entity, $\mathcal{E}$ and $\mathcal{R}$ are the sets of all entities and relations, respectively, in $\mathcal{G}$; and $\mathcal{T}$ is the current time step. We define \emph{learning from scratch} as follows:
\begin{definition}
\textbf{Learning from Scratch.} Given the KG $\mathcal{G}^{\mathcal{T}}$ at time step $\mathcal{T}$, learning from scratch uses a KG embedding method to learn the embeddings of all entities and relations.
\end{definition}

At time step $\mathcal{T}+1$, $\mathcal{G}^{\mathcal{T}}$ becomes $\mathcal{G}^{\mathcal{T}+1}$ with an update involving addition and deletion of triples. The update is not limited to existing entities and relations, and may introduce emerging ones. Here, we define \emph{online learning} as follows:
\begin{definition}
\textbf{Online Learning.} Given the KGs $\mathcal{G}^{\mathcal{T}}$ and $\mathcal{G}^{\mathcal{T}+1}$ at time step $\mathcal{T}+1$ as well as intermediate embedding results at time step $\mathcal{T}$, online learning efficiently learns new embeddings of entities and relations without retraining the whole updated KG $\mathcal{G}^{\mathcal{T}+1}$.
\end{definition}

\begin{figure}[t]
  \centering
  \includegraphics[width=0.85\textwidth]{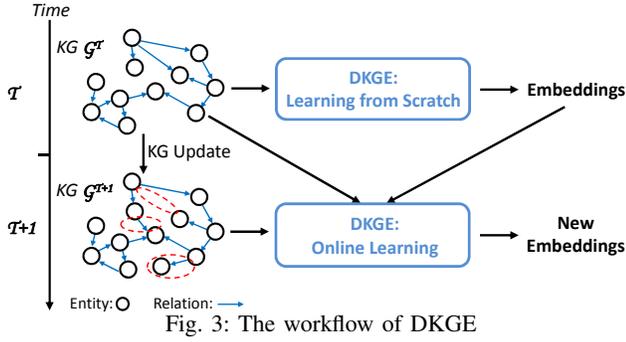}
\caption{The workflow of DKGE.}
\label{fig:dkge}
\end{figure}

Figure~\ref{fig:dkge} illustrates the workflow of our proposed dynamic KG embedding method DKGE, including learning from scratch and online learning.

\section{Learning from Scratch in DKGE}\label{sec:lfs}
In this section, we present the details of learning from scratch in DKGE. The key idea behind DKGE is to preserve $\boldsymbol{h^\star}+\boldsymbol{r^\star}\approx\boldsymbol{t^\star}$ on each triple $(h,r,t)$ in the given KG, where $\boldsymbol{h^\star}$, $\boldsymbol{r^\star}$, and $\boldsymbol{t^\star}$ are the joint embeddings of the head entity, relation, and tail entity incorporating respective contextual information. Such contextual information can provide rich structural features and help well model 1-to-N, N-to-1, and N-to-N relations (discussed in Section~\ref{sec:intro}), thereby enabling DKGE to generate high-quality KG embedding. Thus, we first introduce a unified solution to encode the contexts of entities and relations as vector representations. Then, we describe our strategy to integrate knowledge embeddings of entities and relations with the vector representations of their corresponding contexts. Finally, we define a scoring function and a loss function based on translation operations for parameter training.

\subsection{Context Encoding}\label{subsec:cone}
For entities, the most intuitive context is their neighbor entities. To preserve the structural information among the given entity and its neighbor entities, we define the context of each entity as an undirected subgraph consisting of its neighbor entities and itself. To effectively limit the complexity of DKGE, the number of neighbor entities should not be too large, e.g., only one-hop neighbor entities are considered, but more distant neighbor entities may provide useful information for DKGE, so there is a trade-off between effectiveness and efficiency here. Actually, in our experiments, after using more distant neighbor entities in addition to one-hop ones, it will take much more time for model training, but DKGE's accuracy in link prediction will not be significantly improved (details will be discussed in Section~\ref{subsec:p_sensi}). This is mainly because the farther away neighbor entities are from the given entity, the less relevance they have~\cite{khan2013nema,jin2017querying}, and less relevant neighbor entities may introduce not only useful information but also noise in DKGE. Therefore, we finally only choose one-hop neighbor entities to build the context of each given entity.
\begin{figure}[t]
\centering
\subfigure[] {\includegraphics[width=0.303342732\textwidth]
{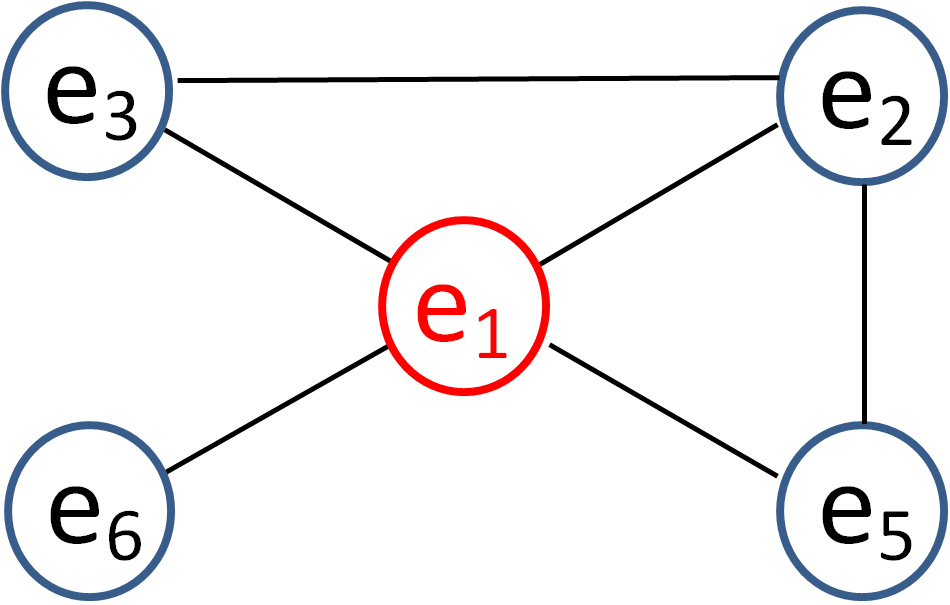}}
\hspace{0.5in}
\subfigure[] {\includegraphics[width=0.29862756\textwidth]
{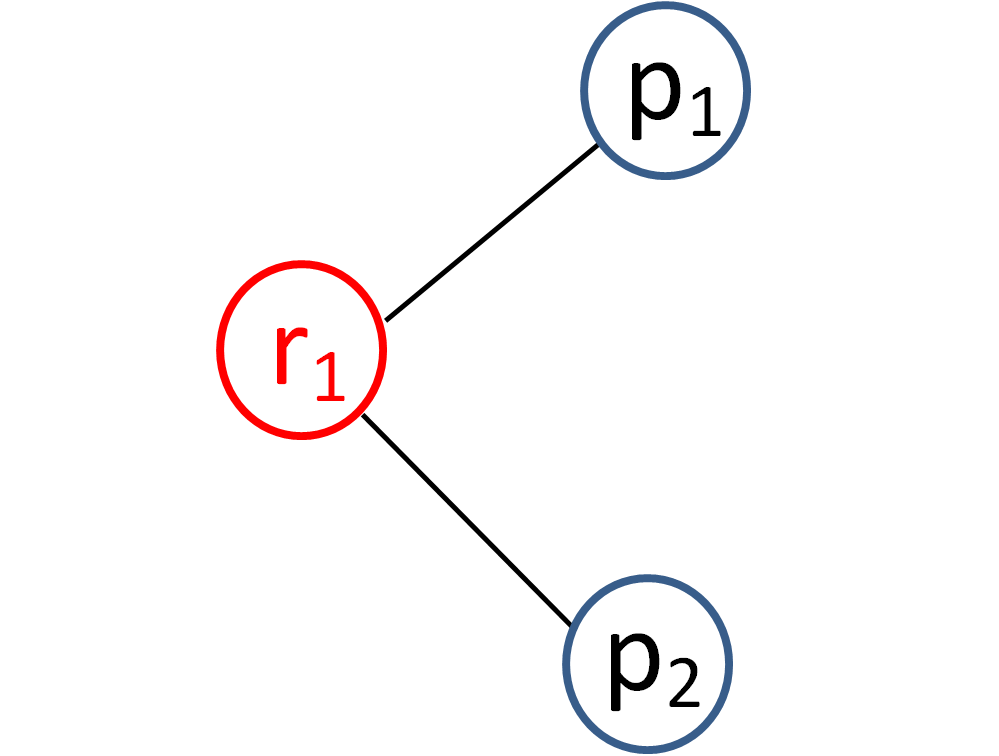}}
\caption{(a) The context of entity $e_1$ in the KG $\mathcal{G}$ at time step $\mathcal{T}+1$ (shown in Figure~\ref{fig:exp_dy}(a)). (b) The context of relation $r_1$ in $\mathcal{G}$ at time step $\mathcal{T}+1$ (also shown in Figure~\ref{fig:exp_dy}(a)). $p_1=(r_1,r_2)$ is a relation path composed of relations $r_1$ and $r_2$, and $p_2=(r_5,r_{4})$ is composed of relations $r_5$ and $r_{4}$.}
\label{fig:exp_con}
\end{figure}

\begin{example}
Figure~\ref{fig:exp_con}(a) shows the context $sg(e_1)$ of entity $e_1$ in the KG $\mathcal{G}$ at time step $\mathcal{T}+1$ given in Figure~\ref{fig:exp_dy}(a). The subgraph $sg(e_1)$ contains the one-hop neighbor entities $\{e_2,e_3,e_5,e_6\}$ and $e_1$ itself. $sg(e_1)$ preserves not only the edges between $e_1$ and its one-hop neighbor entities, but also the edges between such neighbor entities.
\end{example}

Unlike entities, each relation occurs many times in a KG, so it is difficult to choose reasonable neighbor entities or relations as a part of the context of each relation. Here, we choose to use the relation paths connecting the same entity pairs (in the same direction) with each given relation as a part of its context. Then, to capture the structural associations of such relations and relation paths, we transform each relation and its corresponding relation paths connecting the same entity pairs as vertices, and add undirected edges between two vertices if their corresponding relations or relation paths connect the same entity pairs. As a result, we also construct an undirected subgraph as the context of each relation. Similar to the selection of the neighbor entities for each entity's context, to maintain the efficiency of DKGE, we hope that the number of relevant relation paths of a given relation is not too large, so we choose to constrain the length of each relation path. In our experiments, if we consider the relation paths with lengths greater than two, DKGE's accuracy in link prediction will not be significantly improved, but it will cause much more training time (details will be given in Section~\ref{subsec:p_sensi}). Hence, the length of each relation path is constrained as one or two.

\begin{example}
In Figure~\ref{fig:exp_dy}(a), relation $r_1$ and relation path $p_1=(r_1,r_{2})$ are used to link entity $e_1$ to entity $e_5$. Relation $r_1$ and relation path $p_2=(r_5,r_{4})$ are used to link entity $e_1$ to entity $e_2$. Thus, $p_1$ and $p_2$ are neighbor vertices of $r_1$ in the subgraph $sg(r_1)$ (see Figure~\ref{fig:exp_con}(b)), i.e., the context of $r_1$.
\end{example}

Most existing models only assign one representation for each entity and each relation, which is insufficient for online embedding learning after a KG update with addition and deletion of triples, because a revision on the representations of few entities or relations may spread to the entire graph due to the correlations among entities and relations defined in the scoring function. Different from them, each entity or relation in DKGE corresponds to two different representations, i.e., \emph{knowledge embedding} and \emph{contextual element embedding}, which are defined as follows:
\begin{definition}
\textbf{Knowledge Embedding.} When we use a vector representation to denote the given entity $e$ (or, the relation $r$) itself in DKGE, this vector representation is knowledge embedding $\boldsymbol{e}^{k}$ (or, $\boldsymbol{r}^{k}$).
\end{definition}
\begin{definition}
\textbf{Contextual Element Embedding.} When an entity $e$ (or, a relation $r$) denotes a part of the context of other entities or relations in DKGE, its corresponding representation for this role is contextual element embedding $\boldsymbol{e}^{c}$ (or, $\boldsymbol{r}^{c}$).
\end{definition}

Such a setting enables DKGE to perform online learning without re-training the whole KG, which will be introduced in detail in Section~\ref{sec:ol}. Note that the vector representation for the context of an entity or a relation is called \emph{contextual subgraph embedding}, which is defined as follows:
\begin{definition}
\textbf{Contextual Subgraph Embedding.} The context of each entity $e$ (or, relation $r$) is represented as a subgraph $sg(e)$ (or, $sg(r)$), and its vector representation is contextual subgraph embedding $\boldsymbol{sg}(e)$ (or, $\boldsymbol{sg}(r)$), which is formed by combining contextual element embeddings of the entities (or, relations) in the subgraph.
\end{definition}
\begin{figure*}[t]
  \centering
  \includegraphics[width=1\textwidth]{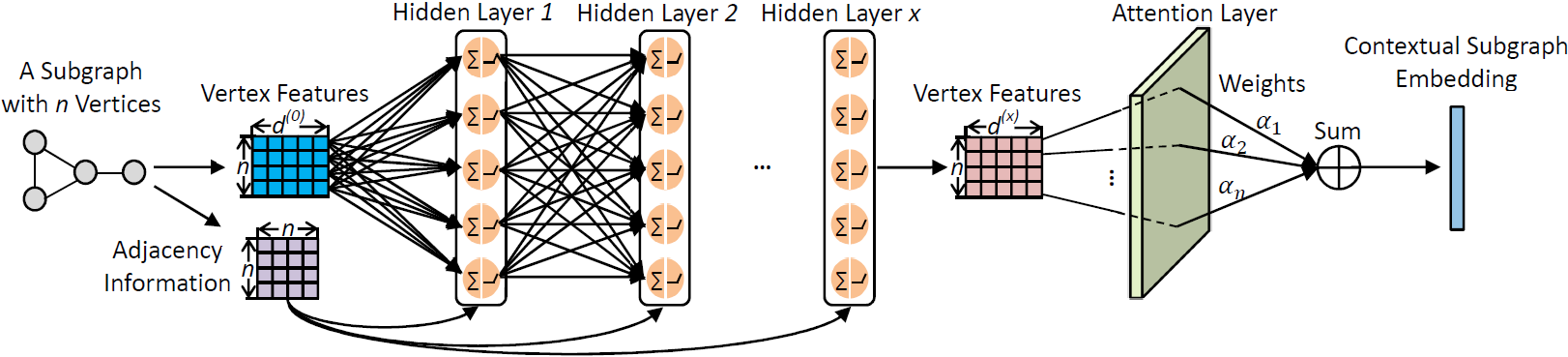}
\caption{The AGCN model. The input is initial vertex features and adjacency information of the given subgraph. Hidden layers conduct convolutional operations to generate new vertex features. The attention layer computes the weight of each vertex. The output contextual subgraph embedding is the weighted sum of all vertices' features.}
\label{fig:agcn}
\end{figure*}

\spara{Why use an attentive GCN?} Since the contexts of entities and relations are all represented as subgraphs, the problem of context encoding is converted into subgraph encoding. Recently, different graph convolutional networks~\cite{bruna2013spectral,henaff2015deep,defferrard2016convolutional,kipf2017semi} have been proposed for feature extraction on arbitrary graphs for machine learning, and have achieved very promising results. The input of a graph convolutional network (GCN) is the initial feature vectors of vertices and the graph structure (i.e., the adjacency matrix). The GCN learns a function of features on the input graph and outputs trained feature vectors of vertices by incorporating neighborhood information, which can capture rich structural information in the input graph. Since our target is to encode a subgraph as a vector, we can use a GCN to learn vectors of all vertices in the input subgraph, and combine them to acquire the vector representation of the subgraph, i.e., contextual subgraph embedding. However, in our scenario, a subgraph is the context of some object (referring to an entity or a relation), so some vertices may be important to this object and some may be useless. Thus, we propose a new attentive GCN model that can automatically assign a weight (i.e., importance) to each vertex for the final combination.

\spara{The Attentive GCN (AGCN) Model.} Figure~\ref{fig:agcn} shows the framework of the AGCN model. Given an object $o$ (an entity or a relation) and its context, i.e., a subgraph with $n$ vertices $\{v_i\}_{i=1}^{n}$, we first build the adjacency matrix $A\in\mathbb{R}^{n\times n}$ and initialize the vertex feature matrix $H^{(0)}\in \mathbb{R}^{n\times d^0}$ with the strategy introduced in Section~\ref{subsec:ptr} ($d^0$ is the number of the initialized features for each vertex). Each row in $H^{(0)}$ is denoted as $\boldsymbol{v_i}$. If $o$ is an entity, then $v_i$ is an entity and $\boldsymbol{v_i}$ is its contextual element embedding. When $o$ is a relation, if $v_i$ is a relation, then $\boldsymbol{v_i}$ denotes its contextual element embedding; if $v_i$ is a relation path consisting of two relations, then $\boldsymbol{v_i}$ is the sum of the contextual element embeddings of these two relations.

Then, we input $H^{(0)}$ and $A$ to the hidden layers to generate the new vertex features incorporating neighborhood information. We apply the propagation rule proposed in~\cite{kipf2017semi} to compute the vertex feature matrix $H^{(\ell)}\in \mathbb{R}^{n\times d^{\ell}}$ ($d^{\ell}$ is the number of features output by the $\ell$th hidden layer) output by the $\ell$th hidden layer with a convolution operation as:
\begin{equation}\label{eq:agcn}
H^{(\ell)}=ReLU(\hat{D}^{-\frac{1}{2}}\hat{A}\hat{D}^{-\frac{1}{2}}H^{(\ell-1)}W^{(\ell)})
\end{equation}
where $ReLU(\cdot)=\max{(0,\cdot)}$ is an activation function, $\hat{A}=A+I$, $I$ is the identity matrix, $\hat{D}$ is the diagonal degree matrix of $\hat{A}$, and $W^{(\ell)}\in \mathbb{R}^{d^{(\ell-1)}\times d^{(\ell)}}$ is the weight matrix of the $\ell$th hidden layer in the AGCN model. The number of hidden layers $x$ means that the AGCN performs $x$ propagation steps during the forward pass and convolves the information from all neighbor vertices up to $x$ hops away. Each entity or relation only has one-hop neighbor vertices in its context, but the neighbor vertices themselves may have two-hop neighbors (see Figure~\ref{fig:exp_con}(a)). Hence, the AGCN used in our scenario contains two hidden layers at most, i.e., $x\in\{1,2\}$. Besides, since each $\boldsymbol{v_i}$ in $H^{(x)}$ may be taken as the input of the AGCN in online learning, we simply set the size of the weight matrix in each hidden layer as $d\times d$, and let $d^0=d^{x}=d$.

The output of the last hidden layer (i.e., $H^{(x)}$) is the input of an attention layer, which computes the weight $\alpha_i(o)$ of each vertex $v_i$ for the object $o$ based on the attention mechanism~\cite{bahdanau2015neural} as follows:
\begin{equation}\label{eq:attscore}
score(\boldsymbol{v_i},\boldsymbol{o}^k)=\boldsymbol{u}^TReLU(\boldsymbol{v_i}\odot\boldsymbol{o}^k)
\end{equation}
\begin{equation}
\alpha_i(o)=\frac{exp(score(\boldsymbol{v_i},\boldsymbol{o}^k))}{\sum_{i=1}^{n}exp(score(\boldsymbol{v_i},\boldsymbol{o}^k))}
\end{equation}
where $\boldsymbol{v_i}\in \mathbb{R}^d$ is a row in $H^{(x)}$, $\boldsymbol{o}^k\in \mathbb{R}^d$ denotes the knowledge embedding of $o$ (initialized by the strategy introduced in Section~\ref{subsec:ptr}), $\boldsymbol{u}\in \mathbb{R}^{d}$ is a parameter vector for the attention layer, $\odot$ means element-wise multiplication, and $score(\boldsymbol{v_i},\boldsymbol{o}^k)$ measures the relevance between $\boldsymbol{v_i}$ and $\boldsymbol{o}^k$.

Finally, we compute the contextual subgraph embedding $\boldsymbol{sg}(o)$ of the subgraph $sg(o)$ by a weighted sum of the vectors of all vertices $\{v_i\}_{i=1}^{n}$ as follows:
\begin{equation}
\boldsymbol{sg}(o)=\sum_{i=1}^{n}\alpha_i(o)\boldsymbol{v_i}
\end{equation}

In summary, our unified solution to context encoding extracts the contexts of entities and relations as subgraphs, and uses an AGCN model to acquire contextual subgraph embeddings. Unlike existing GCNs that operate on a whole big graph, we leverage the small subgraphs of entities to train an AGCN, and the small graphs of relations to train another AGCN.

\subsection{Representation Integration}
After obtaining contextual subgraph embeddings of entities and relations, we integrate them with the knowledge embeddings of entities and relations, to build the joint representation $\boldsymbol{o^\star}$ of each object in the KG. The simplest method is the mean operation, which directly averages the knowledge embedding and contextual subgraph embedding of $o$ to get $\boldsymbol{o^\star}$. The benefit is that we do not need to train any parameter that makes DKGE efficient, but setting the knowledge embedding and contextual subgraph embedding to share the same weight is unreasonable. Another option is the weighting operation, which assigns different weights to the knowledge embedding and contextual subgraph embedding of $o$, but a fixed weight on all dimensions is also inappropriate. Thus, we apply a gate strategy~\cite{xu2017kg} to representation integration; in this strategy, we can assign different weights to different dimensions of a vector as follows:
\begin{equation}\label{eq:gate}
\boldsymbol{o}^{\star}=\boldsymbol{g}\odot\boldsymbol{o}^k+(\boldsymbol{1}-\boldsymbol{g})\odot\boldsymbol{sg}(o)
\end{equation}
where $o$ is an entity or a relation, $\boldsymbol{o}^k$ is its knowledge embedding, $\boldsymbol{sg}(o)$ is its contextual subgraph embedding, $\boldsymbol{g}=logistic(\tilde{\boldsymbol{g}})$ constrains that the value of each element in the gate vector $\boldsymbol{g}$ is in $[0,1]$, and $\tilde{\boldsymbol{g}}\in \mathbb{R}^d$ is a parameter vector. All entities share a $\boldsymbol{g}$ denoted as $\boldsymbol{g}^e$, and all relations share another $\boldsymbol{g}$ denoted as $\boldsymbol{g}^r$.

\subsection{Parameter Training}\label{subsec:ptr}
Since we aim to preserve $\boldsymbol{h^\star}+\boldsymbol{r^\star}\approx\boldsymbol{t^\star}$ on each triple $(h,r,t)$, we define a scoring function as follows:
\begin{equation}\label{eq:score}
f(h,r,t)=\|\boldsymbol{h}^{\star}+\boldsymbol{r}^{\star}-\boldsymbol{t}^{\star}\|_{\ell_{1}}
\end{equation}
where $\boldsymbol{h}^{\star}$, $\boldsymbol{r}^{\star}$ and $\boldsymbol{t}^{\star}$ are computed by Eq.~(\ref{eq:gate}), and $\|\cdot\|_{\ell_{1}}$ denotes the $\ell_{1}$ norm. As discussed earlier in Section~\ref{sec:intro}, Figure~\ref{fig:score_f} shows the architecture of learning embeddings in DKGE. In learning from scratch, we need to train two AGCNs, two gate vectors, and knowledge embeddings as well as contextual element embeddings of all entities and relations. Before training, we first initialize the knowledge embeddings and contextual element embeddings of all entities and relations following the uniform distribution $U(-\frac{6}{\sqrt{d}},\frac{6}{\sqrt{d}})$ (also used in TransE~\cite{bordes2013translating}), where $d$ is the embedding size. The initialized contextual element embeddings form each input initial vertex feature matrix (i.e., $H^{(0)}$ in Eq.~(\ref{eq:agcn})) in our AGCNs.

For training, a margin-based loss function is defined as:
\begin{equation}\label{eq:loss}
\mathcal{L}=\sum\limits_{(h,r,t)\in S}\sum\limits_{(h^{\prime},r,t^{\prime})\in S^{\prime}}\max(0, f(h,r,t)+\gamma-f(h^{\prime},r,t^{\prime}))
\end{equation}where $\gamma$ is the margin, $S$ is the set of correct triples and $S^{\prime}$ is the set of incorrect triples. Since a KG only contains correct triples, we corrupt them by replacing head or tail entities to build $S^{\prime}$. The replacement relies on the techniques of negative sampling. Although there exist some complex negative sampling methods~\cite{cai2018kbgan,wang2018incorporating,zhang2019nscaching} which can effectively improve the quality of KG embedding, we apply a basic negative sampling strategy called Bernoulli sampling~\cite{wang2014knowledge}, which is the most widely used in KG embedding models. We generate an incorrect triple for each correct triple. During training, all parameters, including embeddings, are updated using stochastic gradient descent (SGD) in each minibatch.

\section{Online Learning in DKGE}\label{sec:ol}
In this section, we first introduce our online learning algorithm, and then conduct complexity analysis.

\subsection{Online Learning Algorithm}
Knowledge is not static and always evolves over the time, so KGs should be updated very frequently with addition and deletion of triples. To adapt to such changes, KG embedding should also be dynamically updated in a short time. This requirement raises challenges to existing models as they have to be re-trained on the whole KG with a high time cost. Thus, it is important to build an online embedding learning algorithm which can efficiently generate new high-quality KG embedding based on the results of existing KG embedding.

When the KG has an update, a good online learning algorithm should not only rapidly learn the embeddings of emerging entities and relations, but also consider the impacts on the embeddings of existing entities and relations. Such impacts should be limited in certain regions, not in the entire graph. Based on these principles, we apply the idea of inductive learning so that
\begin{itemize}
   \item parameters in two learnt AGCNs are kept unchanged;
   \item two learnt gate vectors are kept unchanged;
   \item contextual element embeddings of existing entities and relations are kept unchanged.
\end{itemize}

After a KG update, in many triples, the contexts of all entities and relations are unchanged. With unchanged context element embeddings and unchanged parameters in the learnt AGCNs, the contextual subgraph embeddings of such entities and relations are unchanged. Consequently, with unchanged gated vectors and their existing knowledge embeddings, these triples already have $\boldsymbol{h^\star}+\boldsymbol{r^\star}\approx\boldsymbol{t^\star}$, so we also constrain that:
\begin{itemize}
   \item knowledge embeddings of existing entities and relations are kept unchanged as long as their contexts are unchanged.
\end{itemize}

Thus, we only need to learn the knowledge embeddings and contextual element embeddings of emerging entities and relations, as well as the knowledge embeddings of existing entities and relations with changed contexts. This greatly reduces the number of triples which need to be retrained while preserving $\boldsymbol{h^\star}+\boldsymbol{r^\star}\approx\boldsymbol{t^\star}$ on the whole KG.

\begin{figure}[t]
  \centering
  \includegraphics[width=0.9\textwidth]{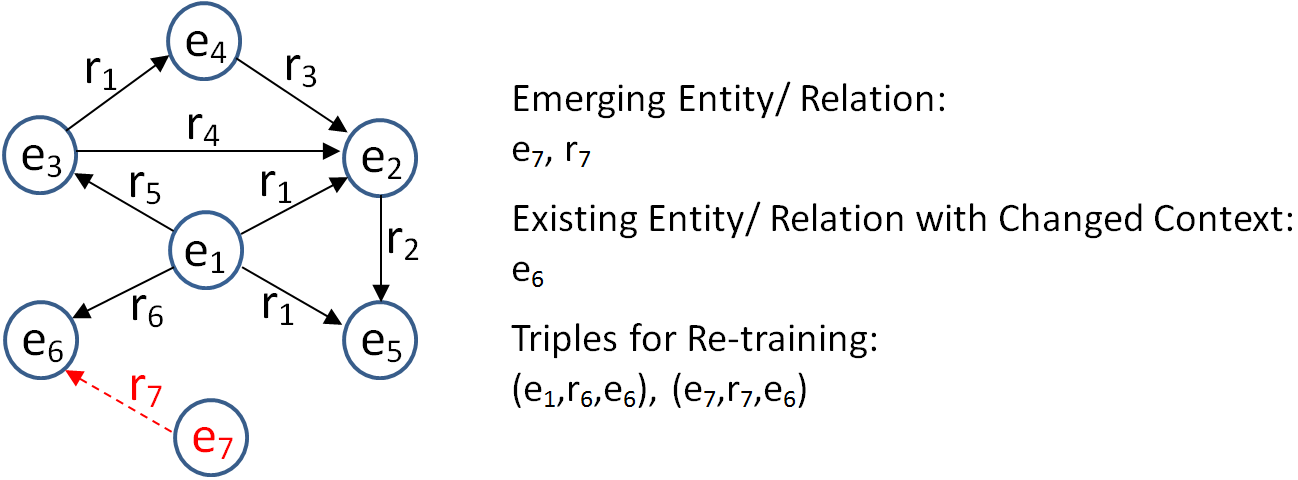}
\caption{The KG $\mathcal{G}$ at time step $\mathcal{T}+2$ ($\mathcal{G}$ at time step $\mathcal{T}+1$ is shown in Figure~\ref{fig:exp_dy}(a)) with the addition of the triple $(e_7,r_7,e_6)$.}
\label{fig:exp_ol}
\end{figure}

\begin{example}
In Figure~\ref{fig:exp_ol}, after adding the triple $(e_7,r_7,e_6)$ into the KG $\mathcal{G}$, we have an emerging entity $e_7$, an emerging relation $r_7$, and one existing entity with changed context $e_6$ (because $e_7$ is a new one-hop neighbor entity of $e_6$). Based on the above idea of online learning, we only need to retrain two triples containing $e_7$, $r_7$, and $e_6$ (i.e., $(e_1,r_6,e_6)$ and $(e_7,r_7,e_6)$) instead of all nine triples in $\mathcal{G}$.
\end{example}

For online learning at time step $\mathcal{T}+1$, the embedding initialization differs from that of learning from scratch. We randomly initialize the knowledge embeddings and contextual element embeddings of emerging entities (relations) following the uniform distribution $U(-\frac{6}{\sqrt{d}},\frac{6}{\sqrt{d}})$. The knowledge embeddings and contextual element embeddings of existing entities (relations) use the embedding results at time step $\mathcal{T}$.

Algorithm~\ref{alg:olearn} shows the whole process of our online learning. Here, we use a 3-tuple $ET=(\mathcal{E}^{\mathcal{T}}\cup\mathcal{R}^{\mathcal{T}},V^k,V^c)$ to record knowledge embeddings and contextual element embeddings of entities and relations, where $\mathcal{E}^{\mathcal{T}}$ and $\mathcal{R}^{\mathcal{T}}$ are the set of entities and relations, respectively, at time step $\mathcal{T}$; $V^k$ is a set of knowledge embeddings, $V^c$ is a set of contextual element embeddings, and each entity or relation corresponds to a knowledge embedding and a contextual element embedding. Given KGs $\mathcal{G}^{\mathcal{T}}$ and $\mathcal{G}^{\mathcal{T}+1}$ at time step $\mathcal{T}$ and $\mathcal{T}+1$, respectively, we first remove the deleted objects (i.e., entities and relations) and their embeddings in $ET$ (\emph{line 3-4}). Then, we add emerging objects and their initialized embeddings into $ET$, and collect all triples containing emerging objects (\emph{line 5-8}). Besides, we collect the triples, each of which has at least one object with changed context (\emph{line 9-13}). Then, we use SGD on the collected triples with the loss function defined in Eq.~(\ref{eq:loss}) to only update knowledge embeddings and contextual element embeddings of emerging entities and relations, as well as knowledge embeddings of existing entities and relations with changed contexts (\emph{line 14-27}). The algorithm will stop based on the performance on a validation set composed of accurate triples. These triples are randomly selected from the given KG, and do not belong to the input of this algorithm. All entities and relations in these triples should occur in other triples used for embedding learning. Finally, our algorithm outputs the updated $ET$ (\emph{line 28}).
\begin{algorithm}[htbp]
\small
\caption{Online Learning}\label{alg:olearn}
\KwIn{KG $\mathcal{G}^{\mathcal{T}}$, entity set $\mathcal{E}^{\mathcal{T}}$, relation set $\mathcal{R}^{\mathcal{T}}$, embedding tuple $ET=(\mathcal{E}^{\mathcal{T}}\cup\mathcal{R}^{\mathcal{T}},V^k,V^c)$ at time step ${\mathcal{T}}$; KG $\mathcal{G}^{\mathcal{T}+1}$, entity set $\mathcal{E}^{\mathcal{T}+1}$, relation set $\mathcal{R}^{\mathcal{T}+1}$ at time step $\mathcal{T}+1$; size of minibatch $b$, learning rate $\lambda$, dimension of embeddings $d$.}
\KwOut{Updated $ET$ at time step $\mathcal{T}+1$.}
      $\Delta \mathcal{E}^d=\mathcal{E}^{\mathcal{T}}-\mathcal{E}^{\mathcal{T}+1}$, $\Delta \mathcal{R}^d=\mathcal{R}^{\mathcal{T}}-\mathcal{R}^{\mathcal{T}+1}$;\\
      $\Delta \mathcal{E}^a=\mathcal{E}^{\mathcal{T}+1}-\mathcal{E}^{\mathcal{T}}$, $\Delta \mathcal{R}^a=\mathcal{R}^{\mathcal{T}+1}-\mathcal{R}^{\mathcal{T}}$;\\
      \ForEach{\textup{object} $o\in \Delta \mathcal{E}^d\cup\Delta \mathcal{R}^d$}
      {
             Remove $o$, its knowledge embedding $\boldsymbol{o}^k$ and contextual element embedding $\boldsymbol{o}^c$ in the embedding tuple $ET$;
      }
      $T^{ol}=\varnothing$;\Comment{\emph{initialize a triple set}}\\
      \ForEach{\textup{object} $o\in \Delta \mathcal{E}^a\cup\Delta \mathcal{R}^a$}
      {
             Add $o$, its knowledge embedding $\boldsymbol{o}^k$ and contextual element embedding $\boldsymbol{o}^c$ into $ET$, and initialize $\boldsymbol{o}^k$ and $\boldsymbol{o}^c$ following the uniform distribution $U(-\frac{6}{\sqrt{d}},\frac{6}{\sqrt{d}})$;\\
             Add all triples in $\mathcal{G}^{\mathcal{T}+1}$ containing $o$ into $T^{ol}$;
      }
      $O^e=\varnothing$;\Comment{\emph{initialize an object set}}\\
      \ForEach{\textup{object} $o\in (\mathcal{E}^{\mathcal{T}+1}-\Delta \mathcal{E}^a)\cup(\mathcal{R}^{\mathcal{T}+1}-\Delta \mathcal{R}^a)$}
      {
             \If{$Context^{\mathcal{T}+1}(o)\neq Context^{\mathcal{T}}(o)$}
             {
                 $O^e=O^e\cup \{o\}$;\\
                 Add all triples in $\mathcal{G}^{\mathcal{T}+1}$ containing $o$ into $T^{ol}$;
             }
      }
      \Loop{}
      {
            $S_{batch}=sample(T^{ol},b)$;\Comment{\emph{sample a minibatch: size b}}\\
            $T_{batch}=\varnothing$;\Comment{\emph{initialize a set of pairs of triples}}\\

            \ForEach{\textup{triple} $(h,r,t)\in S_{batch}$}
            {
                Sample a corrupt triple $(h^{\prime},r,t^{\prime}), h^{\prime}, t^{\prime}\in \mathcal{E}^{\mathcal{T}+1}$;\\
                $T_{batch}=T_{batch}\cup\{(h,r,t),(h^{\prime},r,t^{\prime})\}$;
            }
            \ForEach(\Comment{$h,r,t,h^{\prime},t^{\prime}$}){\textup{object} $o$ \textup{in} $T_{batch}$}
            {
                \If{$o\in \Delta \mathcal{E}^a\cup\Delta \mathcal{R}^a$}
                {
                    $\boldsymbol{o}^k=\boldsymbol{o}^k-\lambda\frac{\partial\mathcal{L}}{\partial\boldsymbol{o}^k}$;\Comment{$\mathcal{L}$\emph{: total loss on} $T_{batch}$}\\
                    $\boldsymbol{o}^c=\boldsymbol{o}^c-\lambda\frac{\partial\mathcal{L}}{\partial\boldsymbol{o}^c}$;\\
                    Update $\boldsymbol{o}^k$ and $\boldsymbol{o}^c$ in $ET$;
                }
                \ElseIf{$o\in O^e$}
                {
                    $\boldsymbol{o}^k=\boldsymbol{o}^k-\lambda\frac{\partial\mathcal{L}}{\partial\boldsymbol{o}^k}$;\\
                    Update $\boldsymbol{o}^k$ in $ET$;
                }
            }
      }
      \Return{$ET$;}
\end{algorithm}
\subsection{Complexity Analysis}
In DKGE, online learning and learning from scratch actually follow the same architecture (shown in Figure~\ref{fig:score_f}), and the difference is that online learning has much fewer triples to train and fewer parameters to update. We analyze the space complexity and time complexity of DKGE in this subsection.

\spara{Space Complexity.}
Given a KG consisting of $|\mathcal{E}|$ entities and $|\mathcal{R}|$ relations, we define the size of the adjacency matrix in the AGCN for entities as $n_e\times n_e$ and that in the AGCN for relations as $n_r\times n_r$. Since the contexts (i.e., subgraphs) of entities (or relations) have different numbers of vertices, to capture all adjacency information of the contexts for entities (or relations), $n_e$ (or $n_r$) should at least be equal to the maximum number of vertices $m_e$ (or $m_r$) among these contexts. However, the KG is dynamic, so the number of vertices in each context may increase, and $n_e$ (or $n_r$) should be larger than $m_e$ (or $m_r$). In our experiments, the maximum number of vertices among the contexts of more than 95\%\footnote{To limit computational resources, similar to~\cite{hamaguchi2017knowledge}, we randomly sample 35 vertices for the remaining 5\% entities and relations to build the contexts, which makes the online updates inexpensive.} of the entities and relations in our datasets is $35$, and we apply zero padding to keep the size of the adjacency matrix of each entity (or relation) as $n_e\times n_e$ (or $n_r\times n_r$). The maximum value of $n_e$ (or $n_r$) is set as $40$ in our experiments. In total, we have $|\mathcal{E}|$ $n_e\times n_e$ adjacency matrices for entities and $|\mathcal{R}|$ $n_r\times n_r$ adjacency matrices for relations, which have the space complexity $O(|\mathcal{E}|n_e^2+|\mathcal{R}|n_r^2)$.

Suppose the AGCNs for entities and relations have $x_e$ and $x_r$ hidden layers, respectively ($x_e$, $x_r\in\{1,2\}$ which are analyzed in Section~\ref{subsec:cone}), since the size of the weight matrix in each hidden layer is set as $d\times d$ (also discussed in Section~\ref{subsec:cone}), we totally have $(x_e+x_r)$ $d\times d$ weight matrices (each hidden layer corresponds to a weight matrix) requiring $O(x_ed^2+x_rd^2)$ space. Besides, the AGCNs for entities and relations respectively have a $d$-dimensional parameter vector (i.e., $\boldsymbol{u}$ in Eq.~(\ref{eq:attscore})) in the attention layer, and this requires $O(2d)$ space. In the gate strategy, all entities (or relations) also correspond to a $d$-dimensional parameter vector (i.e., $\boldsymbol{g}$ in Eq.~(\ref{eq:gate})), respectively, so this part needs $O(2d)$. In addition, each entity or relation has two vector representations, i.e., knowledge embedding and contextual element embedding, so we totally have $(2|\mathcal{E}|+2|\mathcal{R}|)$ $d$-dimensional vectors to represent entities and relations. In summary, online learning and learning from scratch in DKGE share the same space complexity $O(|\mathcal{E}|n_e^2+|\mathcal{R}|n_r^2+(x_e+x_r)d^2+(4+2|\mathcal{E}|+2|\mathcal{R}|)d)$, which is not low because the space complexity is sacrificed to some extend due to the lower time complexity of online learning (introduced later).

\spara{Time Complexity.}
For learning from scratch and online learning, we analyze the time complexities of updating parameters. In learning from scratch, given a KG with $|T|$ triples and the size of a minibatch $b$, we have $\lceil\frac{|T|}{b}\rceil$ minibatches. Suppose each minibatch has $N_e^b$ entities and $N_r^b$ relations on average, so updating their knowledge embeddings requires $O(N_e^bd+N_r^bd)$ time, where $d$ is the dimension of the embedding space. Suppose there are $N_e^c$ entities and $N_r^c$ relations on average composing the contexts of all entities and relations in each minibatch, so updating their contextual element embeddings requires $O(N_e^cd+N_r^cd)$ time. Besides, we need to update the parameters in two AGCNs and the gate strategy. In the AGCN for entities, there are $x_e$ $d\times d$ weight matrices, where $x_e$ is the number of hidden layers, and a $d$-dimensional parameter vector in the attention layer, so updating them in a minibatch requires $O(d+x_ed^2)$ time. Similarly, in the AGCN for relations, updating parameters in a minibatch requires $O(d+x_rd^2)$ time, where $x_r$ is the number of hidden layers. For the gate strategy, updating two $d$-dimensional gate vectors in a minibatch requires $O(2d)$. Thus, for learning from scratch, the total time complexity of updating parameters is $O(\mu\lceil\frac{|T|}{b}\rceil((x_e+x_r)d^2+(N_e^b+N_r^b+N_e^c+N_r^c+4)d))$, where $\mu$ is the number of epochs (one epoch means working through all triples once) when learning from scratch converges.

In online learning, all parameters in two AGCNs and the gate strategy are unchanged, and we only update the knowledge embeddings and context element embeddings of emerging entities and relations, as well as the knowledge embeddings of existing entities and relations with changed contexts. Suppose only $|T^\prime|$ triples, each of which contains at least one emerging object (i.e., entity or relation) or existing object with changed context, need to be retrained, and the size of a minibatch is also $b$, so we have $\lceil\frac{|T^\prime|}{b}\rceil$ minibatches. In each minibatch, on average, suppose there are $N_e^{b\ast}$ existing entities with changed contexts, $N_r^{b\ast}$ existing relations with changed contexts, $N_e^{b\prime}$ emerging entities, and $N_r^{b\prime}$ emerging relations, so updating their knowledge embeddings requires $(N_e^{b\ast}+N_r^{b\ast}+N_e^{b\prime}+N_r^{b\prime})d$ time, where $d$ is the dimension of the embedding space. Suppose there are $N_e^{c\prime}$ emerging entities and $N_r^{c\prime}$ emerging relations on average composing the contexts of all entities and relations in each minibatch, updating their contextual element embeddings requires $(N_e^{c\prime}+N_r^{c\prime})d$ time. Hence, for online learning, the total time complexity of updating parameters is $O(\mu^{\prime}\lceil\frac{|T^\prime|}{b}\rceil(N_e^{b\ast}+N_r^{b\ast}+N_e^{b\prime}+N_r^{b\prime}+N_e^{c\prime}+N_r^{c\prime})d)$, where $\mu^\prime$ is the number of epochs when online learning converges.

For the time complexities of learning from scratch and online learning on the same KG, we can find that $|T^\prime|\ll |T|$, $T^\prime\subseteq T$, and online learning does not require the time cost of updating parameters in AGCNs and the gate strategy $O((x_e+x_r)d^2+4d)$. In a minibatch of size $b$ ($b\leq 500$ on all datasets in our experiments after tuning the hyper-parameters), there is not much difference between $N_e^b+N_r^b+N_e^c+N_r^c$ and $N_e^{b\ast}+N_r^{b\ast}+N_e^{b\prime}+N_r^{b\prime}+N_e^{c\prime}+N_r^{c\prime}$. Since $|T^\prime|\ll |T|$, with the same learning rate, compared with learning from scratch, online learning should have a much faster convergence speed, and usually $\mu^\prime < \mu$. In our experiments, $\mu$ is at least twice $\mu^\prime$ when testing DKGE on different datasets. These findings explain why our online learning has high efficiency.

\spara{Remarks.}
Online learning has much fewer parameters to train compared with learning from scratch in DKGE, which causes that online learning has a smaller model capacity to accumulate underfitting errors~\cite{caruana2001overfitting}, i.e., learning from scratch can well minimize the loss with more parameters. We perform an extensive analysis in Section~\ref{subsec:lp} to understand this effect. Actually, we cannot prove that the embeddings learnt by online learning are optimal due to the machine learning nature, but we shall investigate this more theoretically in our future work.

\section{Experimental Results}\label{sec:exp}
In this section, we present experiments to show the effectiveness and efficiency (especially with respect to online learning) of DKGE on the tasks of link prediction and question answering (QA) in a dynamic environment. The main difference between link prediction and QA is that link prediction aims to predict correct triples which do not exist in the KG, but QA with KG embedding techniques expects to use existing triples in the KG to answer questions. We also analyze the robustness of repeated online learning, investigate the sensitivity of the hyper-parameters of DKGE, and test the scalability of our online learning on a large-scale dataset. The codes for DKGE and the baselines are implemented in Python on the deep learning platform PyTorch. All experiments were executed on an NVIDIA TITAN Xp GPU card (12 GB) of a 64 GB, 2.10 GHz Xeon server. We release the codes for DKGE and all datasets at: \url{https://github.com/lienwc/DKGE/}.

\subsection{Experimental Setup}
\spara{Datasets.} Since there is no publicly available benchmark dataset on link prediction and QA on dynamic KGs, we built four new datasets (two for link prediction, one for QA, and one for scalability testing) from real-world KGs. Each dataset contains multiple snapshots, the differences between which are real changes between different versions of a KG.

\textbf{(1) YAGO-3SP.} YAGO~\cite{mahdisoltani2013yago3} is a large-scale KG built from Wikipedia, WordNet, and GeoNames. YAGO (\url{http://yago-knowledge.org/}) has different versions published at different times. We extracted subsets of YAGO2.5, YAGO3, and YAGO3.1 as three snapshots of our YAGO-3SP dataset. YAGO-3SP was designed for link prediction, and we split each snapshot into a training set, a validation set, and a test set. The three snapshots share the same validation set and test set, in which triples are unchanged in these snapshots.

\textbf{(2) IMDB-30SP.} The Internet Movie Database (IMDB) is a KG consisting of the entities of movies, TV series, actors, directors, among others, and their relationships. IMDB provides daily dumps (\url{https://datasets.imdbws.com/}), and we downloaded them each day from January 22 to February 20 in 2019. We extracted 30 snapshots from such dumps to compose our dataset IMDB-30SP. Similar to YAGO-3SP, IMDB-30SP was also designed for link prediction, and we split each snapshot into a training set, a validation set, and a test set. All snapshots share the same validation set and test set.

\textbf{(3) IMDB-13-3SP.} Different from IMDB-30SP, the size of each snapshot in IMDB-13-3SP is much larger. We kept all the triples about the movies and TV series released after 2013 in the IMDB datasets from January 22 to 24, 2019. With these triples, we built three snaphots. Since IMDB-13-3SP was only utilized to test the scalability of our online learning, we only split each snapshot into a training set and a validation set.

\textbf{(4) DBpedia-3SP.} DBpedia~\cite{lehmann2015dbpedia} is a KG constructed from Wikipedia; different versions of this KG (\url{https://wiki.dbpedia.org/develop/datasets/}) were also published at different times. We extracted subsets from DBpedia3.9 and two subsequent versions as three snapshots of our DBpedia-3SP dataset. DBpedia-3SP was used for case studies on QA, so we only split each snapshot into a training set and a validation set.

The details of the above datasets are given in Table~\ref{tb:datasta}. For each dataset, we recorded: 1) the average numbers of entities (\#Entities), edges (\#Edges), and relations (\#Relations) in different snapshots; 2) the average numbers of added triples (\#Add) and deleted triples (\#Del) between snapshots; and 3) the average number of triples in the training sets (\#Train) of different snapshots, and the number of triples in the validation set (\#Validate) and test set (\#Test). Compared with IMDB-13-3SP, the size of each snapshot in YAGO-3SP, IMDB-30SP, and DBpedia-3SP is much smaller but similar to the sizes of widely used benchmark datasets~\cite{zhang2019nscaching,bordes2013translating,conve2018,schlichtkrull2018modeling,lin2015learning,wang2014knowledge} for static KG embedding.
\begin{table*}[t]
\vsmall
\centering
\caption{Details of our datasets.}
\label{tb:datasta}
\begin{tabular}{|l|c|c|c|c|c|c|c|c|}
\hline \textbf{Datasets}&\textbf{\#Entities}&\textbf{\#Edges}&\textbf{\#Relations}&\textbf{\#Add}&\textbf{\#Del}&\textbf{\#Train}&\textbf{\#Valid}&\textbf{\#Test}\\
\hline
\hline \textbf{YAGO-3SP}&27,009&130,757&37&950&150&124,757&3,000&3,000\\
\hline \textbf{IMDB-30SP}&243,148&627,096&14&9,379&2,395&621,096&3,000&3,000\\
\hline \textbf{IMDB-13-3SP}&3,244,455&7,923,773&14&17,472&18,405&7,913,773&10,000&-\\
\hline \textbf{DBpedia-3SP}&66,967&106,211&968&1,005&103&103,211&3,000&-\\
\hline
\end{tabular}
\end{table*}

\spara{Baselines.} We compared our method DKGE with the following baselines in link prediction on YAGO-3SP and IMDB-30SP. \textbf{(1) puTransE}~\cite{tay2017non}: the only existing model supporting online KG embedding learning for dynamic KGs. \textbf{(2) ConvE}~\cite{conve2018}: the state-of-the-art static deep learning based KG embedding model. \textbf{(3) RotatE}~\cite{sun2018rotate}: the state-of-the-art rotation based static KG embedding model. \textbf{(4) ComplEx}~\cite{trouillon2016complex}: in the research of static KG embedding by matching compositions of head-tail entity pairs with their relations, ComplEx is one of the best models in both effectiveness and efficiency. \textbf{(5) TransE}~\cite{bordes2013translating}: the classic static KG embedding model using translation operations on entities and relations. \textbf{(6) GAKE}~\cite{feng2016gake}: similar to DKGE, the static KG embedding model GAKE simultaneously models triples themselves and graph structural contexts in embedding learning.

We used publicly available codes (implemented in Python on PyTorch) of ConvE, RotatE, ComplEx, and TransE from~\cite{conve2018},~\cite{sun2018rotate},~\cite{trouillon2016complex}, and~\cite{han2018openke}. Since the codes of GAKE (published by the authors) were implemented in C++ and puTransE does not release source codes, we implemented them in Python on PyTorch. For training, we adopted early stopping based on the Hits@10 (introduced in Section~\ref{subsec:lp}) on the validation set, and set the maximum number of epochs as $800$.
\subsection{Link Prediction}\label{subsec:lp}
Link prediction~\cite{wang2017knowledge} in a KG is typically defined as the task of predicting an entity that has a specific relation with another given entity, i.e., predicting the head entity $h$ given the relation $r$ and tail entity $t$ (denoted as $(?,r,t)$), or predicting the tail entity $t$ given the head entity $h$ and relation $r$ (denoted as $(h,r,?)$). Rather than requiring one best result, this task usually ranks a set of candidate entities from the KG.

\spara{Evaluation Metrics.} In the test phase, for each triple $(h,r,t)$ in the test set, we replaced the head entity $h$ (or tail entity $t$) with each entity $e$ in the snapshot to construct a triple $(e,r,t)$ (or $(h,r,e)$), and ranked all $e$ based on the score calculated by the scoring function (e.g., Eq.~(\ref{eq:score}) for DKGE). If a constructed triple occurs in the training set, then the corresponding entity $e$ will not participate in the ranking process, as training data cannot be used in testing. Based on such ranking results, we can get the rank of the original correct entity in each test triple, and we followed the same evaluation metrics of effectiveness used in ConvE~\cite{conve2018} as follows. \textbf{(1) Mean Rank (MR)}: the average rank of all head entities and tail entities in test triples. \textbf{(2) Mean Reciprocal Rank (MRR)}: the average multiplicative inverse of the ranks for all head entities and tail entities in test triples. \textbf{(3) Hits@$\boldsymbol{K}$}: the proportion of ranks not larger than $K$ for all head entities and tail entities in test triples. Besides, in order to evaluate the efficiency of DKGE and baselines, we recorded their training time.

\spara{Hyper-Parameters.} In link prediction on dynamic datasets, we selected optimal hyper-parameters for DKGE and baselines on the first snapshot of each dataset. Each model directly uses such optimal hyper-parameters on subsequent snapshots. In DKGE, the hyper-parameters include embedding size, initial learning rate, size of minibatch, margin, the number of hidden layers of the AGCN for entities, and the number of hidden layers of the AGCN for relations. Given the ranges of each hyper-parameter, we chose the optimal hyper-parameters via grid search according to the Hits@10 on the validation set (details introduced in Section~\ref{subsec:p_sensi}). For ConvE, RotatE, ComplEx, TransE, and GAKE, we applied the same strategy to select the optimal hyper-parameters. puTransE is a non-parametric model without requiring hyper-parameter tuning, so we randomly selected one group of hyper-parameters for testing given the ranges of the hyper-parameters. The details of hyper-parameters tuning for baselines will also be given in Section~\ref{subsec:p_sensi}.

\begin{figure}[t]
\centering
\subfigure[YAGO-3SP] {\includegraphics[width=0.45\textwidth]
{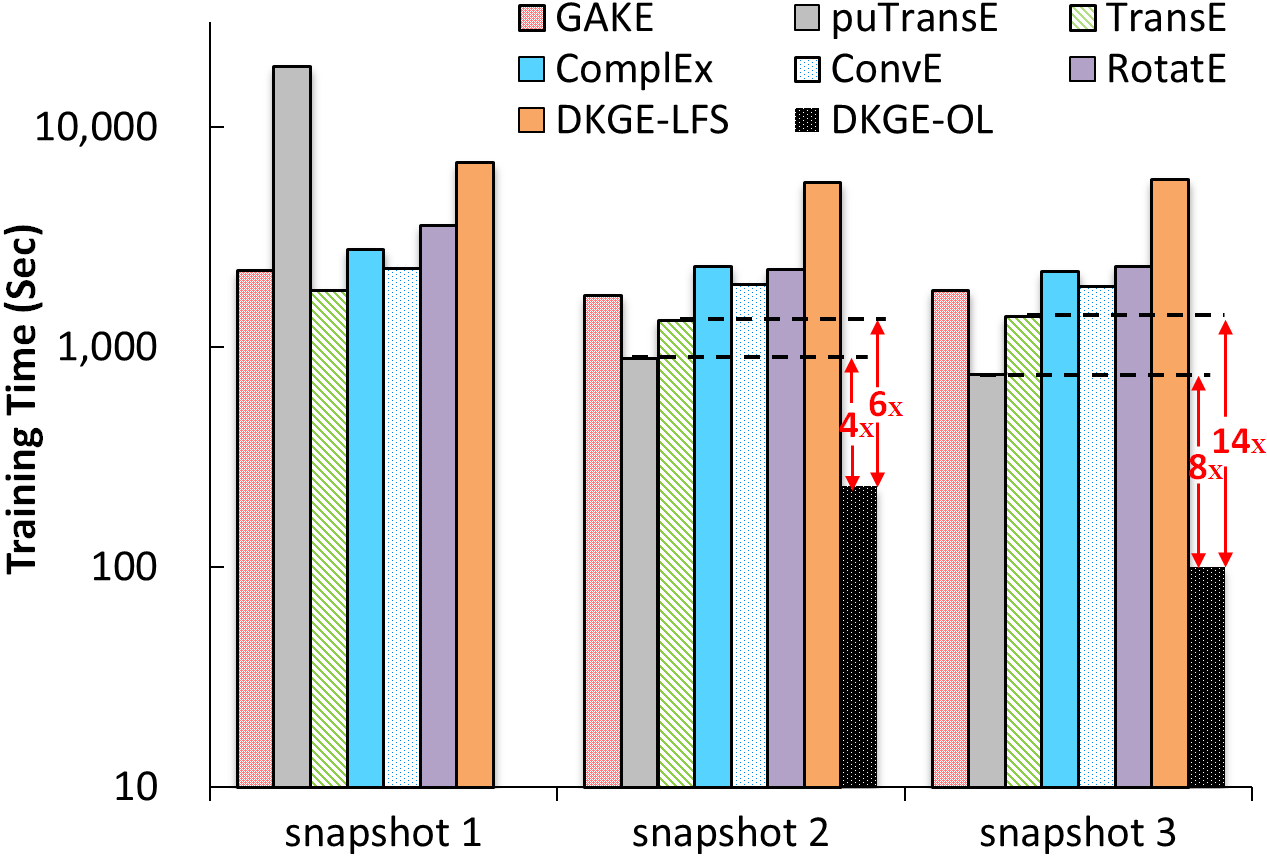}}
\hspace{0.8cm}
\subfigure[IMDB-30SP] {\includegraphics[width=0.45\textwidth]
{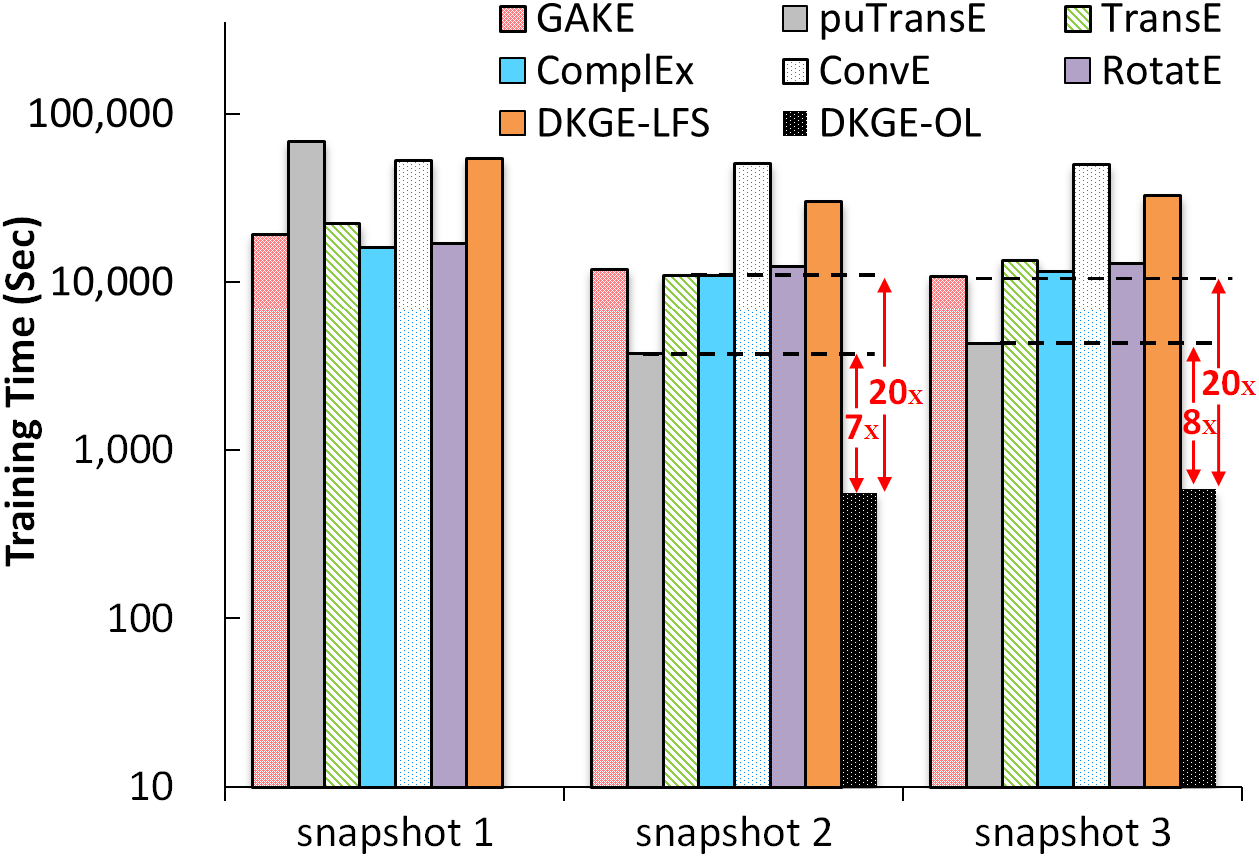}}
\caption{The comparison results on efficiency.}
\label{fig:lp_time}
\end{figure}

\spara{Effectiveness and Efficiency.} We tested DKGE and baselines on all snapshots of YAGO-3SP and the first three snapshots of IMDB-30SP. Given the first snapshot of each dataset, DKGE and baselines train embeddings from scratch on all triples in the training set. When faced with subsequent snapshots, the dynamic KG embedding models DKGE and puTransE can use online learning to acquire new embeddings, but other static KG embedding baselines can only be retrained on all triples in the training set. The comparison results on effectiveness and efficiency between DKGE and the baselines are shown in Table~\ref{tb:com_lp} and Figure~\ref{fig:lp_time}, respectively. On the second and third snapshots in two datasets, note that we tested both learning from scratch and online learning in DKGE, but for puTransE, we only tested its online learning.
\begin{table*}[t]
\tiny
\centering
\caption{The comparison results on effectiveness (our methods: DKGE-LFS (learning from scratch) and DKGE-OL (online learning)).}
\label{tb:com_lp}
\begin{tabular}{|l|l||c|c|c|c|c||c|c|c|c|c|}
\hline
\multicolumn{2}{|c||}{\multirow{2}*{}}&\multicolumn{5}{c||}{YAGO-3SP}&\multicolumn{5}{c|}{IMDB-30SP}\\
\cline{3-12}
\multicolumn{2}{|c||}{}&MR&MRR&Hits@10&Hits@3&Hits@1&MR&MRR&Hits@10&Hits@3&Hits@1\\
\hline
\hline
\multirow{7}*{Snapshot 1}&GAKE&2,984&0.150&0.237&0.155&0.098&5,798&0.116&0.213&0.119&0.081\\
 &puTransE&938&0.180&0.262&0.188&0.130&3,518&0.122&0.188&0.132&0.096\\
 &TransE&666&0.348&0.508&0.385&0.263&2,443&0.330&0.499&0.368&0.242\\
 &ComplEx&1,155&0.412&0.532&0.451&0.342&5,671&0.285&0.454&0.315&0.200\\
 &ConvE&1,614&0.450&0.525&0.473&0.402&6,713&0.271&0.412&0.317&0.208\\
 &RotatE&\bf{446}&\bf{0.463}&\bf{0.555}&\bf{0.480}&\bf{0.403}&\bf{1,087}&\bf{0.380}&\bf{0.587}&\bf{0.427}&\bf{0.283}\\
 &\bf{DKGE-LFS}&\bf{643}&\bf{0.460}&\bf{0.545}&\bf{0.479}&\bf{0.411}&\bf{2,390}&\bf{0.381}&\bf{0.569}&\bf{0.431}&\bf{0.283}\\
\hline
\hline
\multirow{8}*{Snapshot 2}&GAKE&3,012&0.141&0.218&0.151&0.095&5,542&0.116&0.218&0.118&0.079\\
 &puTransE&897&0.186&0.259&0.195&0.133&3,506&0.119&0.182&0.134&0.092\\
 &TransE&975&0.300&0.460&0.340&0.226&2,415&0.323&0.492&0.363&0.235\\
 &ComplEx&995&0.380&0.521&0.420&0.303&6,037&0.274&0.453&0.314&0.184\\
 &ConvE&1,319&\bf{0.450}&0.538&0.473&\bf{0.406}&7,011&0.265&0.418&0.301&0.203\\
 &RotatE&\bf{429}&\bf{0.460}&\bf{0.548}&\bf{0.482}&\bf{0.397}&\bf{1,061}&\bf{0.391}&\bf{0.598}&\bf{0.439}&\bf{0.287}\\
 &\bf{DKGE-LFS}&\bf{723}&0.440&\bf{0.545}&\bf{0.475}&0.393&\bf{2,347}&0.378&\bf{0.570}&0.425&\bf{0.280}\\
 &\bf{DKGE-OL}&749&0.440&0.539&0.473&0.393&2,841&\bf{0.380}&0.567&\bf{0.428}&\bf{0.282}\\
\hline
\hline
\multirow{7}*{Snapshot 3}&GAKE&2,873&0.140&0.220&0.156&0.087&5,623&0.116&0.219&0.116&0.081\\
 &puTransE&1,082&0.173&0.247&0.180&0.130&3,522&0.123&0.187&0.134&0.095\\
 &TransE&959&0.304&0.460&0.335&0.226&2,560&0.326&0.494&0.360&0.242\\
 &ComplEx&974&0.392&0.524&0.426&0.325&5,824&0.267&0.461&0.306&0.172\\
 &ConvE&1,531&\bf{0.447}&0.531&0.470&\bf{0.404}&7,129&0.260&0.422&0.292&0.190\\
 &RotatE&\bf{412}&\bf{0.452}&\bf{0.545}&\bf{0.482}&0.395&\bf{1,103}&\bf{0.391}&\bf{0.598}&\bf{0.440}&\bf{0.285}\\
 &\bf{DKGE-LFS}&\bf{747}&0.445&\bf{0.542}&\bf{0.476}&\bf{0.397}&\bf{2,368}&\bf{0.383}&\bf{0.571}&\bf{0.435}&\bf{0.285}\\
 &\bf{DKGE-OL}&809&0.442&\bf{0.542}&0.473&0.395&2,976&0.377&0.561&0.427&0.281\\
\hline
\end{tabular}
\end{table*}

As Table~\ref{tb:com_lp} shows, when comparing to all baselines on both datasets, we can find that the learning from scratch in DKGE (\textbf{DKGE-LFS}) ranks in the top two in most evaluation metrics, which reflects the superiority of our model. DKGE-LFS and the online learning in DKGE (DKGE-OL) are comparable to the state-of-the-art static KG embedding models, including the rotation-based model and the deep learning based model ConvE, and sometimes are even better. This is mainly because all static KG embedding baselines except GAKE focus on modeling triples themselves, but neglect structural contexts, which can bring useful information in embedding learning, while GAKE models structural contexts, but neglects relational constraints between entities. Compared with the dynamic KG embedding model puTransE, DKGE-LFS and DKGE-OL have much better performance, because DKGE solves two major problems (see Section~\ref{sec:intro}) of puTransE. DKGE-OL and DKGE-LFS have close performance, which also shows the effectiveness of our online learning.

In Figure~\ref{fig:lp_time}, we can see that DKGE-LFS does not have the best efficiency on the first snapshot of each dataset, but when we used DKGE-OL on the second and third snapshots, the training time is much less. Compared with static KG embedding models, the training time of DKGE-OL on YAGO-3SP and IMDB-30SP is at least $6$ and $20$ times faster, respectively. Compared with the online learning of puTransE, the training times of DKGE-OL on YAGO-3SP and IMDB-30SP are at least $4$ and $7$ times faster, respectively. This demonstrates the high efficiency of our model.

\begin{figure}[t]
  \centering
  \includegraphics[width=0.8\textwidth]{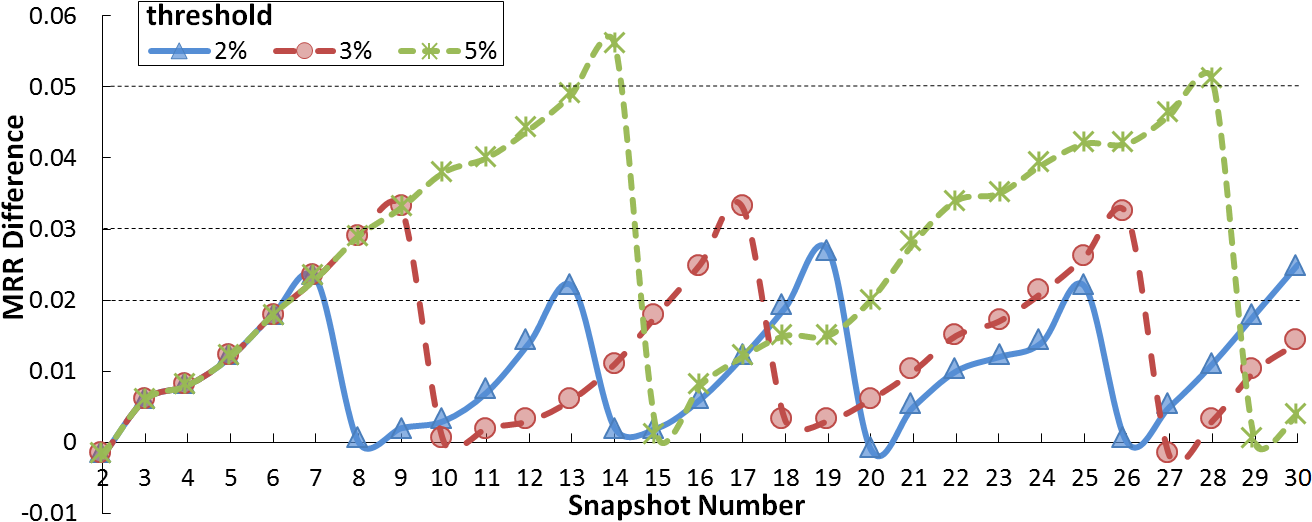}
\caption{The robustness analysis for repeated online learning.}
\label{fig:lp_robust}
\end{figure}

\spara{Robustness w.r.t. Repeated Updates.} DKGE-OL is the online version of DKGE-LFS. The quality of the learnt embeddings may become lower after continuously conducting online learning a number of times. Thus, we performed robustness analysis on IMDB-30SP for DKGE-OL. On the first snapshot, we applied DKGE-LFS with the optimal hyper-parameters. Starting from the second snapshot, we applied DKGE-LFS and DKGE-OL, and recorded their MRR difference, which gets larger as testing more snapshots. When the MRR difference exceeds a threshold on the $y$th snapshot, the embeddings generated by DKGE-LFS will be taken as the input of the DKGE-OL used on the $(y+1)$th snapshot. As a result (see Figure~\ref{fig:lp_robust}), if we set the threshold as $5\%$ (or $3\%$, or $2\%$), we should perform DKGE-LFS after continuously using DKGE-OL $14$ (or $8$, or $6$) days (the IMDB dataset is updated once per day). The MRR difference between DKGE-LFS and DKGE-OL will not increase significantly within a short time period, which indicates the good robustness of our online learning.

\begin{figure}[t]
  \centering
  \includegraphics[width=0.8\textwidth]{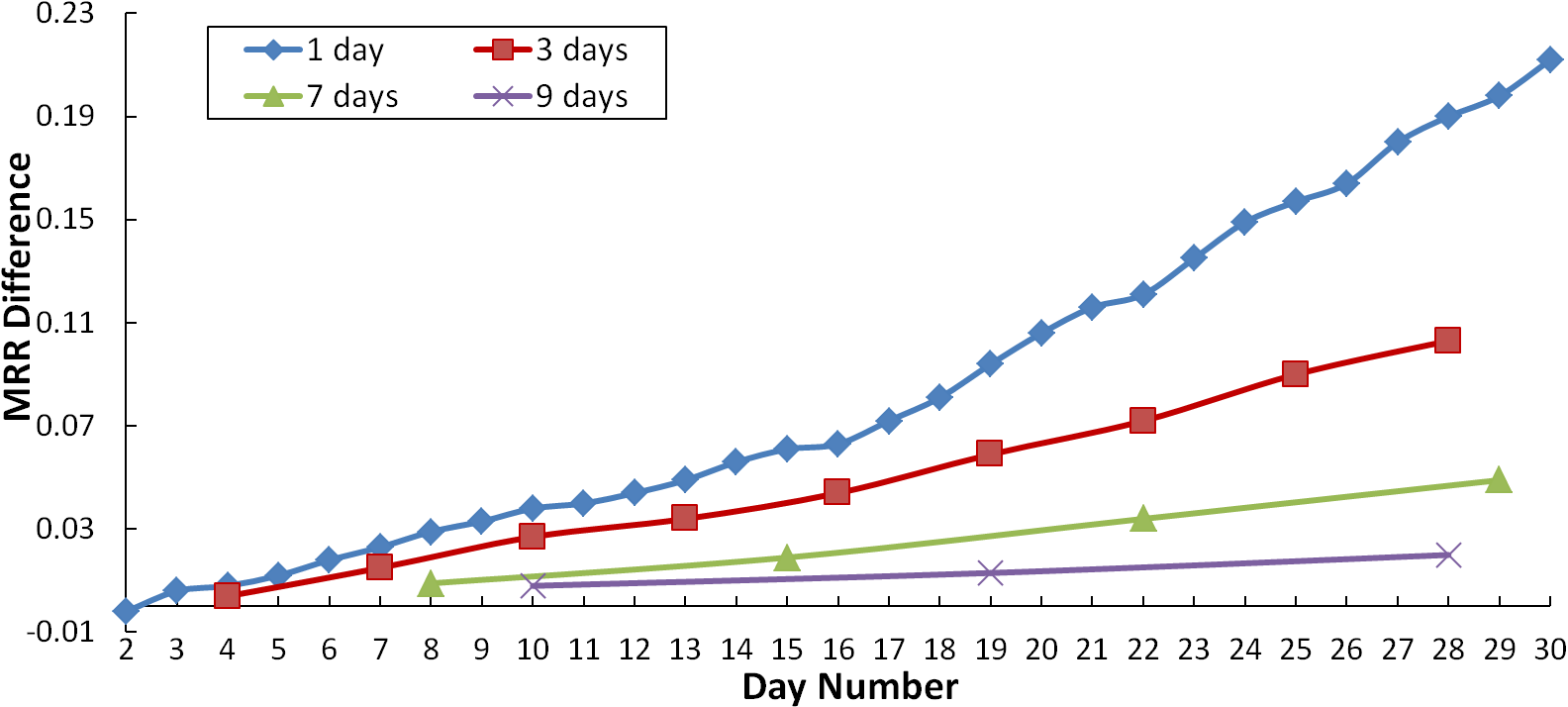}
\caption{The update aggregation analysis for repeated online learning.}
\label{fig:lp_robust_reason}
\end{figure}

We argue that the main reason for the degradation of DKGE-OL is: DKGE-OL has much fewer parameters to train compared with DKGE-LFS, which causes that DKGE-OL has a smaller model capacity to accumulate underfitting errors~\cite{caruana2001overfitting}. We also find that the loss of DKGE-OL is $11\%$ higher on average than that of DKGE-LFS (for $3\%$ triples update on average) on the test set in the above robustness evaluation. To further validate our argument, we aggregated the daily updates of IMDB-30SP once every 3 (or 7, or 9) days, performed DKGE-LFS and DKGE-OL, and recorded their MRR difference. Figure~\ref{fig:lp_robust_reason} shows that aggregating more KG updates for online learning (i.e., more parameters to train) can lower the MRR difference between DKGE-LFS and DKGE-OL. However, training more parameters in DKGE-OL will cost more time, e.g., the time of training DKGE-OL once every 3 days is at least 5 times more than that of training DKGE-OL once per day. Thus, whether to aggregate more KG updates for online learning should be decided by users' own needs.

\subsection{Question Answering}
In this subsection, we conducted case studies on QA in a dynamic environment, and DKGE can help find correct answers without writing structured queries in query languages (e.g., SPARQL). We prepared ten questions (see Table~\ref{tb:qa}), and the answer of each question exists in each snapshot of DBpedia-3SP. The answers of the same question in different snapshots may be different because knowledge is always changing. Here, each question is a simple question which can be denoted in the form of a triple $(h, r, ?)$, and the head entity $h$ and relation $r$ existing in DBpedia-3SP are implied in the question. Thus, similar to the idea of the state-of-the-art KG embedding based QA system~\cite{huang2019knowledge}, we identified the head entity and relation expressed by each question. This step was manually finished, and it can also be solved by automatic strategies, such as the template-based method~\cite{zheng2018question} or the learning-based model~\cite{huang2019knowledge}. For example, we denoted the question \ding{172} in Table~\ref{tb:qa} as (\emph{Kobe\_Bryant}, \emph{draftTeam}, \emph{?}). For embedding learning, given the first snapshot of DBpedia-3SP, we utilized DKGE-LFS to train KG embedding. The process of choosing hyper-parameters will be introduced in Section~\ref{subsec:p_sensi}. Given the second and third snapshots, we applied DKGE-OL with the selected hyper-parameters to generate new KG embedding. Finally, based on the identified head entity and relation of each question, as well as all parameters in DKGE including their embeddings, we inferred the tail entity as the answer by ranking all entities in DBpedia-3SP using the score calculated by the scoring function (i.e., Eq.~(\ref{eq:score})) in DKGE.
\begin{table*}[t]
\tiny
\centering
\caption{The prepared questions and their answers in DBpedia-3SP.}
\label{tb:qa}
\begin{tabular}{|c|c|c|c|c|}
\hline \multirow{2}*{}&\multirow{2}*{\textbf{Question}}&\multicolumn{3}{|c|}{\textbf{Answer in}}\\
\cline{3-5}&&\textbf{Snapshot 1}&\textbf{Snapshot 2}&\textbf{Snapshot 3}\\
\hline
\hline \ding{172}&\it{Which team drafts Kobe Bryant?}&New\_Orleans\_Hornets&Charlotte\_Hornets&Charlotte\_Hornets\\
\hline \ding{173}&\it{Who is the chief of China's Central Military Commision?}&Hu\_Jintao&Xi\_Jinping&Xi\_Jinping\\
\hline \ding{174}&\it{Which team does Dwight Howard play for?}&Los\_Angeles\_Lakers&Houston\_Rockets&Houston\_Rockets\\
\hline \ding{175}&\it{Who is the coach of Golden State Warriors?}&Mark\_Jackson\_(basketball)&Steve\_Kerr&Steve\_Kerr\\
\hline \ding{176}&\it{Who has the most caps of Portugal national football team?}&Lu\'is\_Figo&Cristiano\_Ronaldo&Cristiano\_Ronaldo\\
\hline \ding{177}&\it{Who is the top scorer of Argentina national football team?}&Gabriel\_Batistuta&Gabriel\_Batistuta&Lionel\_Messi\\
\hline \ding{178}&\it{Which team does Luke Walton coach?}&Golden\_State\_Warriors&Golden\_State\_Warriors&Los\_Angeles\_Lakers\\
\hline \ding{179}&\it{Which team does Byron Scott coach?}&Cleveland\_Cavaliers&Los\_Angeles\_Lakers&Los\_Angeles\_Lakers\\
\hline \ding{180}&\it{Which team does Kevin Garnett play for?}&Boston\_Celtics&Brooklyn\_Nets&Brooklyn\_Nets\\
\hline \ding{181}&\it{Who is the wife of Martin Fowler in EastEnders?}&Sonia\_Fowler&Sonia\_Fowler&Stacey\_Slater\\
\hline
\end{tabular}
\end{table*}

\spara{Evaluation Metrics.} To evaluate the effectiveness of DKGE for QA on the dynamic dataset DBpedia-3SP, we used: \textbf{(1) Mean Rank (MR)}: the average rank of all correct answers in each snapshot; \textbf{(2) Mean Reciprocal Rank (MRR)}: the average multiplicative inverse of ranks for all correct answers in each snapshot; \textbf{(3) P@$\boldsymbol{K}$}: the average proportion of the correct answers (in each snapshot) in top-$K$ ranks. Besides, we recorded the training time of DKGE on each snapshot.

\spara{Results Analysis.} DKGE-LFS takes 5,221 seconds on the first snapshot of DBpedia-3SP to train embeddings. For DKGE-OL on the second and third snapshots, it only takes 595 seconds and 672 seconds, respectively. With the embedding results and identified head entities and relations of questions, we constructed a triple for each question, to solve the QA problem. For example, we constructed a triple (\emph{Kobe\_Bryant}, \emph{draftTeam}, \emph{New\_Orleans\_Hornets}) for question \ding{172} (in Table~\ref{tb:qa}) given the first snapshot. Note that not all constructed triples exist in the KG, i.e., the training set of each snapshot in DBpedia-3SP. We found that all constructed triples of questions \ding{172}-\ding{178} exist in the training sets, but the constructed triples of questions \ding{179}-\ding{181} do not. Table~\ref{tb:qa_res} shows the evaluation results of QA on DBpedia-3SP using DKGE. For all questions, given different snapshots, the performance on the same evaluation metric is good and close, which reflects that DKGE is effective for QA in a dynamic environment. The QA performance on questions \ding{172}-\ding{178} is much better than that of questions \ding{179}-\ding{181}. The reason is that the embeddings of entities and relations for the constructed triples of questions \ding{172}-\ding{178} have been optimized to constrain $\boldsymbol{h^\star}+\boldsymbol{r^\star}\approx\boldsymbol{t^\star}$ during training, but for the constructed triples of questions \ding{179}-\ding{181}, they do not have such optimizations. From another perspective, users cannot query the triples which do not exist in the KG by writing structured queries in query languages, but QA with KG embedding techniques can provide help by embedding calculations, e.g., all correct answers of questions \ding{179}-\ding{181} on different snapshots occur in the top-10 ranks.
\begin{table}[t]
\footnotesize
\centering
\caption{Evaluation results of QA using DKGE}
\label{tb:qa_res}
\begin{tabular}{|l|l|c|c|c|}
\hline
&\textbf{DBpedia-3SP}&\textbf{MR}&\textbf{MRR}&\textbf{P@1}\\
\hline
\hline
\multirow{3}*{\textbf{All Questions}}&Snapshot 1&5&0.497&0.400\\
 &Snapshot 2&5&0.487&0.400\\
 &Snapshot 3&5&0.484&0.400\\
\hline
\multirow{3}*{\textbf{Question \ding{172}-\ding{178}}}&Snapshot 1&4&0.648&0.571\\
 &Snapshot 2&4&0.638&0.571\\
 &Snapshot 3&4&0.640&0.571\\
\hline
\multirow{3}*{\textbf{Question \ding{179}-\ding{181}}}&Snapshot 1&7&0.145&0\\
 &Snapshot 2&8&0.126&0\\
 &Snapshot 3&8&0.120&0\\
\hline
\end{tabular}
\end{table}

\subsection{Parameter Sensitivity}\label{subsec:p_sensi}
This subsection first gives the details of selecting the optimal hyper-parameters (for DKGE and the baselines) on the first snapshot of YAGO-3SP and IMDB-30SP via grid search according to the Hits@10 on the validation set. With the optimal hyper-parameters of DKGE, we only varied one hyper-parameter each time to test its effects in link prediction.

For hyper-parameter tuning, we first set the ranges of the shared hyper-parameters of DKGE and the baselines as follows: embedding size (i.e., dimensionality): $\{20, 30, 50, 80, 100\}$, initial learning rate: $\{0.001, 0.003, 0.005, 0.01, 0.03\}$, and the size of minibatch: $\{100, 200, 300, 400, 500\}$. Then, we set the ranges of specific hyper-parameters belonging to each model based on~\cite{sun2018rotate,conve2018,trouillon2016complex,feng2016gake,bordes2013translating,tay2017non} as follows:
\begin{itemize}
  \item DKGE: margin: $\{1,2,5,8,10\}$, the number of hidden layers of the AGCN for entities: $\{1,2\}$, and the number of hidden layers of the AGCN for relations: $\{1,2\}$;
  \item RotatE: margin: $\{1,2,5,8,10\}$, and self-adversarial sampling temperature: $\{0.5,1\}$;
  \item ConvE: embedding dropout: $\{0,0.1,0.2\}$, projection layer dropout: $\{0,0.1,0.3\}$, feature map dropout: $\{0,0.1,0.2\}$, and label smoothing: $\{0,0.1,0.2\}$;
  \item ComplEx: regularization parameter: $\{0.001,0.01,0.03\}$, and the number of negative samples per positive sample: $\{5,10\}$;
  \item TransE: margin: $\{1,2,5,8,10\}$, and norm: $\{\ell_1,\ell_2\}$;
  \item GAKE: prestiges of neighbor context: $\{1,0.8\}$, path context: $\{0.1,0.2\}$, and edge context: $\{0.1,0.2\}$;
  \item puTransE: margin: $\{1,2,5,8,10\}$, and the number of embedding spaces: $\{500,800,1000,1200\}$.
\end{itemize}

For YAGO-3SP, the optimal hyper-parameters of DKGE are as follows: embedding size: $100$, initial learning rate: $0.005$, size of minibatch: $500$, margin: $10$, the number of hidden layers in the AGCN for entities: $1$, and the number of hidden layers in the AGCN for relations: $1$. For IMDB-30SP, all optimal hyper-parameters of DKGE are the same with the ones used on YAGO-3SP except the size of minibatch, which is $200$.
\begin{figure}[htbp]
\centering
\subfigure[Effects of the embedding size (i.e., dimensionality).]{\includegraphics[width=0.9\textwidth]
{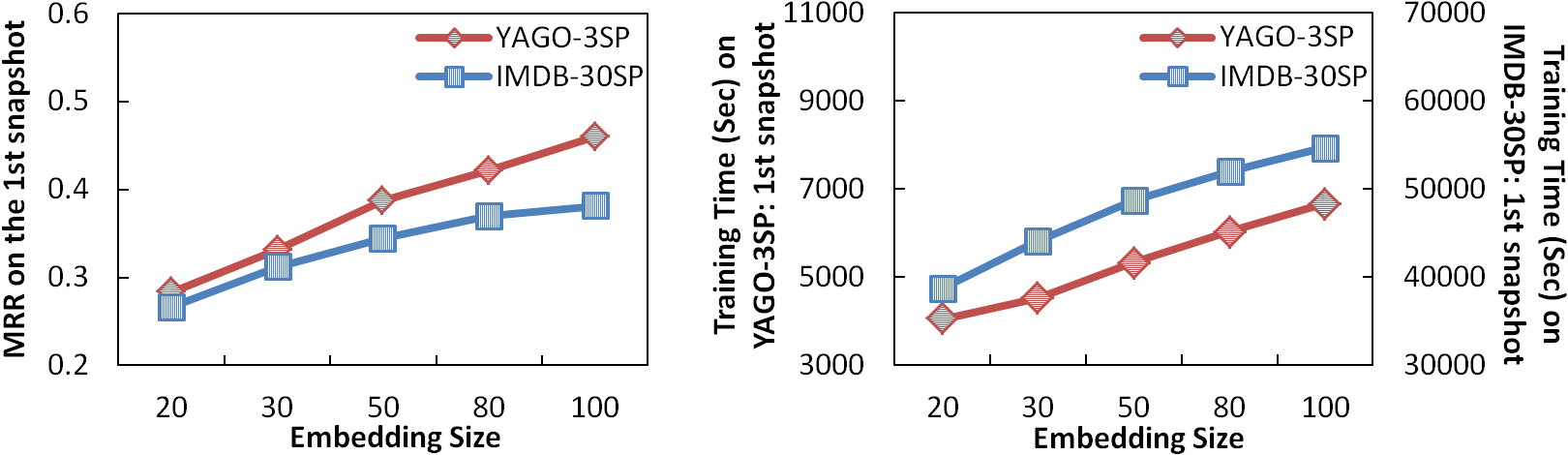}}
\subfigure[Effects of the initial learning rate.] {\includegraphics[width=0.9\textwidth]
{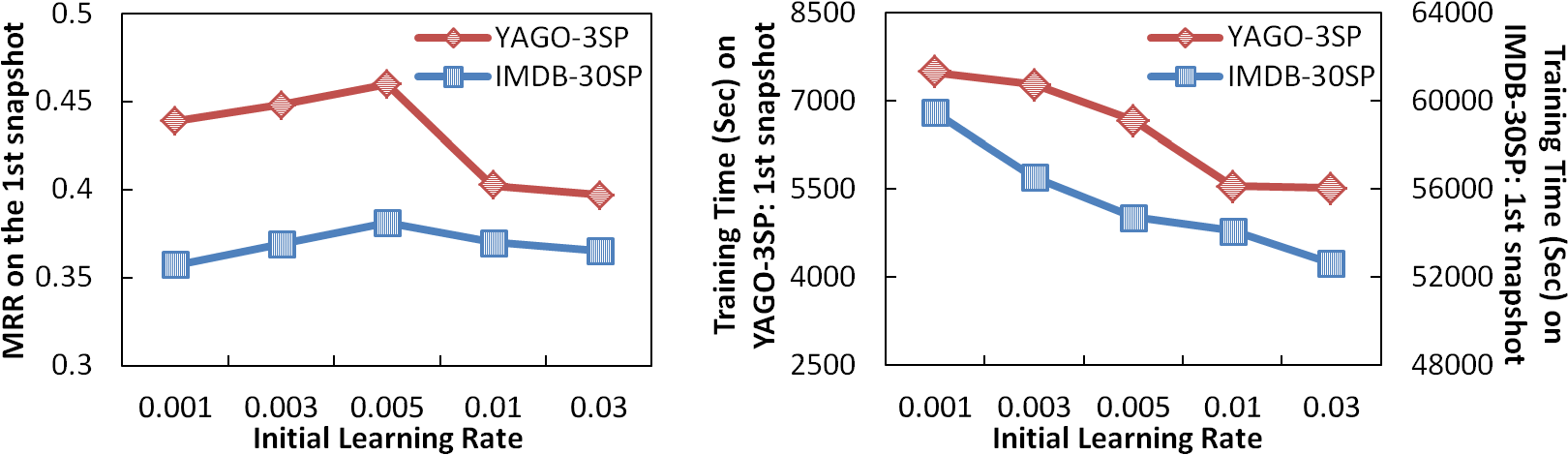}}
\subfigure[Effects of the size of minibatch.] {\includegraphics[width=0.9\textwidth]
{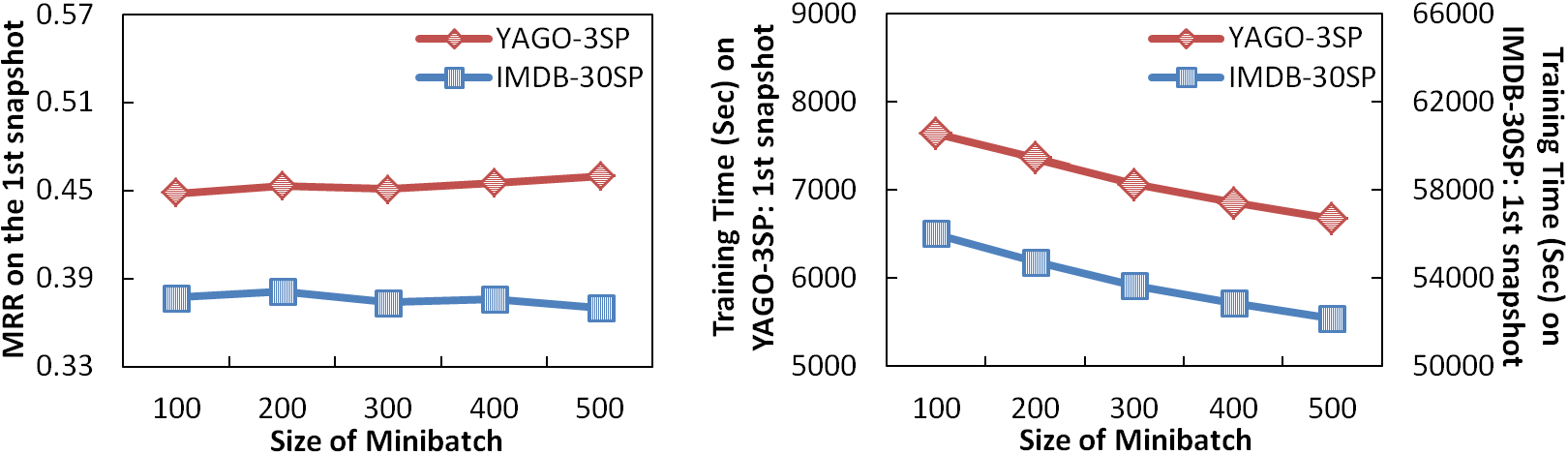}}
\subfigure[Effects of the margin.] {\includegraphics[width=0.9\textwidth]
{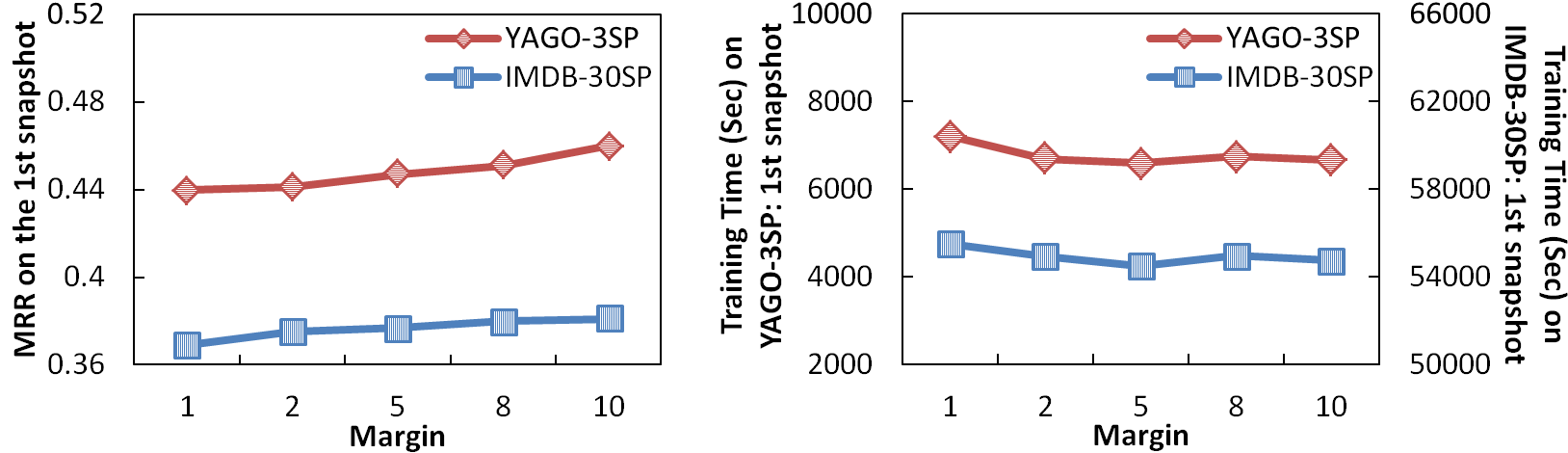}}
\caption{Effects of the embedding size, initial learning rate, size of minibatch, and margin in DKGE for link prediction.}
\label{fig:ps_dkge}
\end{figure}

Figure~\ref{fig:ps_dkge} shows the effects of different hyper-parameters on YAGO-3SP and IMDB-30SP in link prediction. When the embedding size increases, DKGE will have better MRR, but the training time will increase. Similarly, the larger the margin, the better the MRR, but it does not significantly affect the training time. To ensure effectiveness, the initial learning rate should be neither too large nor too small, and the larger initial learning rate, the less training time of DKGE-LFS. The size of minibatch does not significantly affect the effectiveness, but as it increases, the training time will also increase.

Table~\ref{tb:ps_ne_hlayer} shows the effects of the maximum hop of neighbor entities and the number of hidden layers of the AGCN for entities. We can see that if using more distant neighbor entities to build the contexts of entities, the MRR of DKGE in link prediction will not be significantly improved (and may even decrease), but this will cost much more training time, especially for online learning. Adding hidden layers will also not greatly improve the effectiveness. This is why we only consider one-hop neighbor entities in DKGE.

Table~\ref{tb:ps_rpl_hlayer} shows the effects of the maximum length of relation paths and the number of hidden layers of the AGCN for relations. If using the relation paths with the length greater than two to build the contexts of relations, the MRR will not be significantly improved, but the training time will increase considerably in online learning. Since the vertices in the contexts of relations only have one-hop or two-hop neighbors (introduced in Section~\ref{subsec:cone}), we tested the number of hidden layers in $\{1,2\}$, and one hidden layer always achieves the better MRR.
\begin{table}[t]
\footnotesize
\centering
\caption{Effects of the maximum hop of neighbor entities $\alpha$ and number of hidden layers $x_e$ of the AGCN for entities (snapshot 1: learning from scratch, snapshot 2: online learning).}
\label{tb:ps_ne_hlayer}
\begin{tabular}{|l|c|c|c||c|c|c|}
\hline
\multirow{3}*{($\alpha$, $x_e$)}&\multicolumn{3}{c||}{YAGO-3SP}&\multicolumn{3}{c|}{IMDB-30SP}\\
\cline{2-7}
&\multicolumn{2}{c|}{Snapshot 1}&Snapshot 2&\multicolumn{2}{c|}{Snapshot 1}&Snapshot 2\\
\cline{2-7}
&MRR&Time (s)&Time (s)&MRR&Time (s)&Time (s)\\
\hline
\hline
(1, 1)&0.460&\bf{6,861}&\bf{232}&\bf{0.381}&\bf{54,135}&\bf{560}\\
(1, 2)&0.455&7,348&312&0.370&55,231&699\\
(2, 1)&0.460&7,836&651&0.375&57,910&1,251\\
(2, 2)&\bf{0.465}&8,032&704&0.380&58,523&1,642\\
(2, 3)&0.453&8,288&718&0.368&59,127&1,889\\
(3, 2)&0.460&9,018&1,610&0.361&60,907&2,843\\
(3, 3)&0.450&9,229&1,856&0.343&62,378&3044\\
(4, 2)&0.448&10,123&4,501&0.340&66,303&7,630\\
\hline
\end{tabular}
\end{table}
\begin{table}[t]
\footnotesize
\centering
\caption{Effects of the maximum length of relation paths $\beta$ and number of hidden layers $x_r$ of the AGCN for relations (snapshot 1: learning from scratch, snapshot 2: online learning).}
\label{tb:ps_rpl_hlayer}
\begin{tabular}{|l|c|c|c||c|c|c|}
\hline
\multirow{3}*{($\beta$, $x_r$)}&\multicolumn{3}{c||}{YAGO-3SP}&\multicolumn{3}{c|}{IMDB-30SP}\\
\cline{2-7}
&\multicolumn{2}{c|}{Snapshot 1}&Snapshot 2&\multicolumn{2}{c|}{Snapshot 1}&Snapshot 2\\
\cline{2-7}
&MRR&Time (s)&Time (s)&MRR&Time (s)&Time (s)\\
\hline
\hline
(1, 1)&0.412&\bf{6,784}&\bf{176}&0.341&\bf{52,936}&\bf{307}\\
(1, 2)&0.412&7,043&196&0.332&54,770&323\\
(2, 1)&\bf{0.460}&6,861&232&0.381&54,135&560\\
(2, 2)&0.453&7,007&287&0.370&55,002&677\\
(3, 1)&\bf{0.460}&6,988&840&\bf{0.385}&53,895&1,601\\
(3, 2)&0.442&7,285&885&0.372&55,883&1,548\\
(4, 1)&0.450&6,936&1,904&0.381&55,779&3,066\\
\hline
\end{tabular}
\end{table}

Based on the above analysis, to simultaneously ensure the effectiveness and efficiency, we applied the optimal hyper-parameters used on YAGO-3SP to train DBpedia-3SP for QA and IMDB-13-3SP for scalability testing.

\subsection{Scalability}
\begin{figure}[t]
  \centering
  \includegraphics[width=0.7\textwidth]{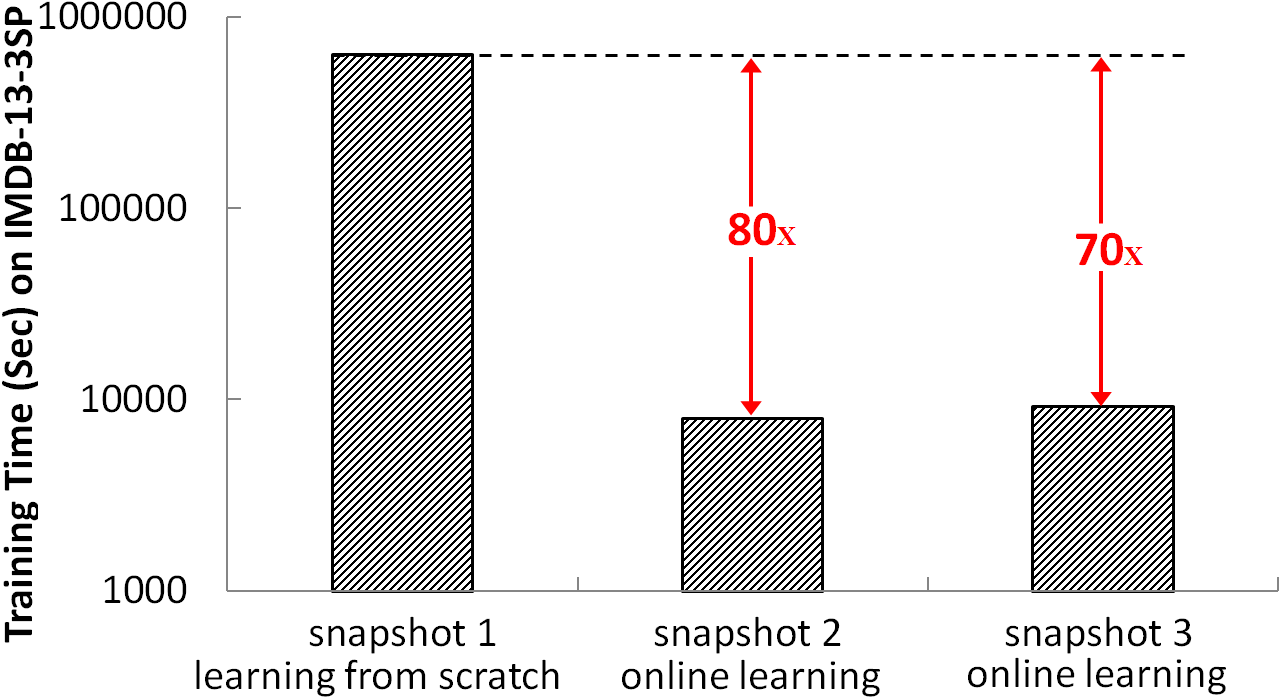}
\caption{The training time of DKGE on IMDB-13-3SP.}
\label{fig:scala_test}
\end{figure}

We tested the scalability of DKGE on a large-scale dataset IMDB-13-3SP. We applied DKGE-LFS to snapshot 1 and DKGE-OL to snapshots 2 and 3. In Figure~\ref{fig:scala_test}, although DKGE-LFS takes approximately one week to train KG embedding, DKGE-OL only needs about two hours to finish training, which means the online learning in DKGE scales well on the real-world large-scale dynamic KG.

\section{Related Work}\label{sec:rewo}
\spara{Static KG Embedding.}
Almost all existing KG embedding models (for a survey, see~\cite{wang2017knowledge}) represent entities and relations in the KG in low-dimensional vector spaces,
and define a scoring function on each triple to measure its plausibility.
This scoring function captures correlations among entities and relations.
By maximizing the total plausibility of all triples in the KG, we obtain embeddings of entities and relations.

A line of research is to use translation operations to model correlations among entities and relations.
The most typical work is TransE~\cite{bordes2013translating} which takes the relation
between entities corresponding to a translation between the embeddings of entities. TransH~\cite{wang2014knowledge} improves TransE by projecting the embedding of each entity into a relation-specific hyperplane, and performing the same translation operations of TransE at this hyperplane.
TransR~\cite{lin2015learning} follows a similar idea of TransH, the only difference is
to replace relation-specific hyperplanes with relation-specific spaces. TransR also has several extensions such as TransD~\cite{ji2015knowledge}
and TranSparse~\cite{ji2016knowledge}.

Another direction of research is to match compositions of head-tail entity pairs with their relations.
The earliest work is RESCAL~\cite{nickel2011three} which represents each triple as a tensor. Each relation is denoted as a matrix modeling pairwise interactions
between entity vectors by a bilinear function. DisMult~\cite{yang2015embedding} simplifies RESCAL by restricting
relation matrices to diagonal matrices, to reduce the number of parameters, but it cannot handle symmetric relations. To solve this problem, ComplEx~\cite{trouillon2016complex} models
KG embedding in a complex space and takes the conjugate of the embedding of each tail
entity before calculating the bilinear map.

Neural networks are employed to produce high-quality KG embedding.
R-GCN~\cite{schlichtkrull2018modeling} is a relational graph convolutional network model which utilizes convolutional
operators on the semantic information in local graph structures to generate KG embedding.
ConvE~\cite{conve2018} is a multi-layer convolutional neural network model, which learns KG embedding by its deep structure and 2D convolutions. Besides, rotation-based static KG embedding models have achieved the state-of-the-art performance in link prediction, and the representative model is RotatE~\cite{sun2018rotate}, which maps
the entities and relations to the complex vector space and defines relations as rotations between entities.

All of the above models only consider triples themselves in embedding learning while neglecting graph structural features, such as neighbor information. To address this issue, GAKE~\cite{feng2016gake} was proposed to embed KGs using co-occurrence probabilities of entities, relations and structural contexts.

Unlike our DKGE, all the aforementioned models only embed static KGs, and cannot support online embedding learning.

\spara{Dynamic KG Embedding.}
The most relevant model to our paper is puTransE~\cite{tay2017non}, which creates multiple parallel embedding spaces from local parts of the given KG, and selects the global highest energy score for link prediction across the embedding spaces.
When facing a KG update, puTransE can train new embedding spaces (for triple addition) and delete existing spaces (for triple deletion) for online learning.
As discussed in Section~\ref{sec:intro}, our method DKGE can solve the two major problems of puTransE to generate high-quality embeddings. iTransA~\cite{jia2018knowledge} supports the online optimization of entity-specific and relation-specific margins, but for embedding learning, it needs to retrain all triples in the KG.

CKGE~\cite{daruna2021continual} applies continual learning in knowledge graph embedding, and assumes that the model observes disjoint subsets of a complete knowledge graph which differs from our setting of online KG embedding. Continual learning~\cite{lesort2020continual} is a machine learning paradigm learning from a sequence of partial experiences where all data are not available at once, when given a potentially unlimited stream of data. However, our online KG embedding can get all triples at the current time step.

There also exist some models~\cite{dasgupta2018hyte,jiang2016towards,trivedi2017know,goel2020diachronic,jin2020recurrent,li2021temporal} on temporal KG embedding~\cite{cai2022temporal}, which aims to incorporate the temporal information of triples into embedding learning, to better perform link prediction and time prediction. They cannot update KG embedding in an online manner.

\spara{Dynamic Graph Embedding.}
Different from KG embedding, graph embedding usually only learns vertex embeddings based on structural proximities without considering relational semantics on edges. Recently, some graph embedding models have focused on dealing with dynamics to acquire high-quality evolving embeddings of vertices. DynamicTriad~\cite{zhou2018dynamic} and DyRep~\cite{trivedi2019dyrep} preserve structural information and evolving patterns of a graph to learn vertex embeddings, which are used for vertex classification, link prediction, and etc. at the next time step. However, DynamicTriad can only be applied when vertices are fixed, and DyRep does not support the online updating of existing vertex embeddings. MTSN~\cite{liu2021motif} learns vertex embeddings by integrating motif features and temporal
evolution information into graph neural networks, but it cannot update vertex embeddings in an online scenario, and it was extended to large scale temporal data for scalable temporal graph embedding~\cite{gao2022scalable}. GraphSAGE~\cite{hamilton2017inductive} is an inductive model utilizing neighbor attributes to generate embeddings for previously unseen data, but it cannot update embeddings of existing vertices when the graph has changed. DepthLGP~\cite{ma2018depthlgp} leverages a Laplacian Gaussian process and deep learning to learn vertex embeddings, and it only infers the embeddings of new vertices when facing a graph update. Both DHPE~\cite{zhu2018high} and DANE~\cite{li2017attributed} use matrix decomposition to learn vertex embeddings of a static graph, and matrix perturbation to incrementally update vertex embeddings to adapt to graph changes. DNE~\cite{du2018dynamic} extends skip-gram based graph embedding methods to the dynamic scenario. It decomposes the skip-gram objective function to support learning the embedding of each vertex separately, so that it can calculate the embeddings of new vertices. It also measures the influence of graph changes on the original vertices to update their embeddings. DRLAN~\cite{liu2020dynamic} models time-evolving structural and attribute changes with matrix calculations to incrementally update vertex embeddings in a timely manner.

Although DHPE, DANE, DNE, and DRLAN can incrementally compute the embeddings of new vertices and update existing vertices' embeddings after a graph update, when we need to learn edge (i.e., relation) embeddings and consider various semantic correlations among vertices and edges in dynamic KG embedding, these models cannot be applied.

\section{Conclusions}\label{sec:con}
In this paper, we presented a context-aware dynamic knowledge graph (KG) embedding method DKGE, which can not only learn embeddings from scratch, but also support online embedding learning. Compared with state-of-the-art static and dynamic KG embedding models on dynamic datasets, DKGE has comparable effectiveness and much better efficiency in online learning. The experimental results also show the value of DKGE for link prediction and question answering in a dynamic environment, and the good robustness and scalability of the online learning in DKGE.

As for the future work, we will theoretically study how to alleviate the problem of accumulated underfitting errors brought by the online learning in DKGE. We will also extend DKGE to continual KG embedding learning especially for privacy-sensitive applications when we only have limited access to the KG. Besides, we plan to improve DKGE in real-world knowledge graph based dynamic question answering.

\section{Acknowledgements}
This work is supported by the NSFC (Grant No. 62006040, U21A20488), the Novo Nordisk Foundation (Grant No. NNF22OC0072415), the Project for the Doctor of Entrepreneurship and Innovation in Jiangsu Province (Grant No. JSSCBS20210126), the Fundamental Research Funds for the Central Universities, and ZhiShan Young Scholar Program of Southeast University.

\bibliographystyle{elsarticle-num}
\bibliography{paper}

%
%
%
%
%
%
%
%
\end{document}